\documentclass{article}

\usepackage{PRIMEarxiv}

\usepackage[utf8]{inputenc} 
\usepackage[T1]{fontenc}    
\usepackage{hyperref}       
\usepackage{url}            
\usepackage{booktabs}       
\usepackage{amsfonts}       
\usepackage{nicefrac}       
\usepackage{microtype}      
\usepackage{lipsum}
\usepackage{amssymb,amsthm,amsmath,bm}
\usepackage{graphicx}
\usepackage{algorithmic}
\usepackage{paralist}
\usepackage{tabularx}
\newcolumntype{Y}{>{\centering\arraybackslash}X}
\usepackage[algo2e,ruled,noend,linesnumbered]{algorithm2e}

\def\x{\mathbf{x}}
\def\y{\mathbf{y}}
\def\H{\mathbf{H}}
\def\I{\mathbf{I}}
\def\K{\mathbf{K}}
\def\R{\mathbf{R}}

\def\bsigma{\bm{\sigma}}
\def\btheta{\bm{\theta}}
\def\bmu{\bm{\mu}}
\def\bSigma{\bm{\Sigma}}

\def\cF{\mathcal{F}}
\def\cN{\mathcal{N}}

\title{BAMCAFE:~A Bayesian Machine Learning Advanced Forecast Ensemble Method for Complex Turbulent Systems with Partial Observations}

\author{
 Nan Chen \\
  Department of Mathematics, \\
  University of Wisconsin-Madison, \\
  Madison, WI 53706, USA \\
  \texttt{chennan@math.wisc.edu} \\
   \And
 Yingda Li \\
  Department of Mathematics, \\
  University of Wisconsin-Madison, \\
  Madison, WI 53706, USA \\
  \texttt{yli678@wisc.edu} \\
}

\begin{document}
\maketitle
\begin{abstract}
Ensemble forecast based on physics-informed models is one of the most widely used forecast algorithms for complex turbulent systems. A major difficulty in such a method is the model error that is ubiquitous in practice. Data-driven machine learning (ML) forecasts can reduce the model error but they often suffer from the partial and noisy observations. In this paper, a simple but effective Bayesian machine learning advanced forecast ensemble (BAMCAFE) method is developed, which combines an available imperfect physics-informed model with data assimilation (DA) to facilitate the ML ensemble forecast. In the BAMCAFE framework, a Bayesian ensemble DA is applied to create the training data of the ML model, which reduces the intrinsic error in the imperfect physics-informed model simulations and provides the training data of the unobserved variables. Then a generalized DA is employed for the initialization of the ML ensemble forecast. In addition to forecasting the optimal point-wise value, the BAMCAFE also provides an effective approach of quantifying the forecast uncertainty utilizing a non-Gaussian probability density function that characterizes the intermittency and extreme events. It is shown using a two-layer Lorenz 96 model that the BAMCAFE method can significant improve the forecasting skill compared to the typical reduced-order imperfect models with bare truncation or stochastic parameterization for both the observed and unobserved large-scale variables. It is also shown via a nonlinear conceptual model that the BAMCAFE leads to a comparable non-Gaussian forecast uncertainty as the perfect model while the associated imperfect physics-informed model suffers from large forecast biases.

\end{abstract}


\section{Introduction}
Forecasting complex turbulent systems is an important task in many areas, particularly in geophysics, engineering, neuroscience, and climate science~\cite{majda2016introduction, strogatz2018nonlinear, wilcox1988multiscale, sheard2009principles}. These systems are often characterized by a large dimensional phase space with strong nonlinear interactions between different spatial and temporal scales, which transfer energy throughout the system. Intermittent instability, extreme events, non-Gaussian probability density functions (PDFs), and multiscale dynamics are typical characteristics in these systems~\cite{farazmand2019extreme, moffatt2021extreme}. Because of such features, errors and  uncertainties from various sources (e.g., initializations, parameters and approximations) may get amplified in the forecast stage, which leads to big challenges in effectively forecasting these complex systems. In addition to predicting the optimal point-wise value, an accurate quantification of the forecast uncertainty of these turbulent systems is an equally important issue.

Ensemble forecast based on physics-informed (or parametric) models is one of the most widely used forecast algorithms for predicting turbulent signals~\cite{palmer2019ecmwf, toth1997ensemble, leutbecher2008ensemble}. Starting from a given initial condition, multiple model simulations are conducted resulting in a forecast ensemble.  However, developing effective ensemble forecast algorithms that  enable a skillful forecast with an accurate uncertainty quantification is very challenging mainly for the following three reasons. First, only partial and noisy observations are available in many real-world situations~\cite{kalnay2003atmospheric, lau2011intraseasonal}. For example, temperature at the sea surface are more accessible than inside the deep ocean from satellite observations, and large-scale components of many geophysical systems are more observable than small-scale components.
Such partial and noisy observations often introduce large biases and uncertainties at the initialization stage, which significantly affect the accuracy of the ensemble forecast as time evolves. 
Second, due to the high dimensionality and the complexity of many turbulent systems, it is computationally expensive for even a single run of the forecast, prohibiting the forecast being repeated multiple times to create the ensemble. Third, one of the major and ubiquitous difficulties in applying the parametric model-based ensemble forecast is the model error~\cite{majda2018model, allen2002model}. The presence of the model error is often due to a lack of the perfect understanding of nature and the inadequate resolution in the models because of the limited computing power, where model reduction techniques and parameterizations are often adopted when developing practical approximate models~\cite{palmer2001nonlinear, phillips2004evaluating, rotstayn2000tuning, mou2020data, mou2021data}.

Purely data-driven approaches, such as machine learning (ML) or other non-parametric models have gained great interest in the past decades~\cite{vlachas2018data, chattopadhyay2020data, chattopadhyay2020superparameterization, scher2018toward, scher2018predicting, weyn2020improving, pathak2018model, rasp2021data, raissi2019physics, beucler2019achieving, chattopadhyay2020deep}. Given a sufficiently large amount of accurate training data, ML models with well-designed architectures can extract key information from the high-dimensional complex systems, which allows them for effective forecasts~\cite{vlachas2018data,chattopadhyay2020data,chattopadhyay2020superparameterization,scher2018toward,scher2018predicting}. It is also worthwhile to notice that once the model has been trained, the computational cost for executing these data-driven models is much cheaper (or even negligible) compared to the cost in a typical numerical solver for the parametric models~\cite{chattopadhyay2020superparameterization}. Popular ML models include the long short-term memory (LSTM) networks~\cite{hochreiter1997long,vlachas2018data}, the convolutional neural networks~\cite{lecun1998gradient,weyn2020improving}, and the echo state networks ~\cite{jaeger2007echo,pathak2018model}. However, these ML models may suffer from the polluted, incomplete, and insufficient training data, which appear in many real applications resulting from the partial and noisy observations~\cite{rasp2021data,chattopadhyay2021towards,farchi2020using,brajard2021combining}. Physics-informed ML models partially solve this issue by enforcing some key physics knowledge (e.g., conservation law) as constraints or including them into the  model architecture design~\cite{raissi2019physics,beucler2019achieving,chattopadhyay2020deep,de2019deep}.

Data assimilation (DA), which combines parametric models with observations, plays a vital role in assisting physics-based model forecast of turbulent systems~\cite{law2015data, ghil1991data, kalnay2003atmospheric}. The main benefits of DA are two-fold. First, DA recovers the states of the unobserved variables. Second, by incorporating the information from observations into the available imperfect model, the model error and the observational noise are simultaneously mitigated. Therefore, DA improves the initialization of both the observed and the unresolved state variables that facilitates the ensemble forecast. However, utilizing partial observations to correct the model error via DA only applies to the initialization stage of the standard physics-based model ensemble forecast; the model error continuously enters into the forecast ensembles as time evolves.
On the other hand, DA can also be incorporated into the ML forecast, where promising results have been shown in various approaches~\cite{chattopadhyay2021towards,tomizawa2020combining,bocquet2020bayesian,farchi2020using,brajard2021combining,chen2020can}. In iterative methods, the ML model is treated as a surrogate model in DA~\cite{brajard2020combining,wikner2021using,bocquet2020bayesian}. However, the so-called cold start problem may arise in this type of approach that leads to the potential numerical instability. In addition, the algorithms may take a long time to converge. One remedy is to add a neural network (NN) as a residual to correct the imperfect knowledge-informed outcomes, which helps reduce the number of cycles as well~\cite{farchi2020using,brajard2021combining}. Note that some methods also attempt to predict the uncertainty  by training an ML model on the error residual~\cite{scher2018predicting}. Other ML forecast approaches involving perturbing the initial conditions are also developed to mimic the traditional ensemble forecast approach. However, as was pointed in~\cite{scherEnsemble}, the uncertainty obtained in these methods can be systematically lower than that in the physics-based numerical prediction models.

This paper develops a simple but effective~\underline{Ba}yesian \underline{M}a\underline{c}hine \underline{L}earning \underline{A}dvanced \underline{F}orecast \underline{E}nsemble (BAMCAFE) method. Assume that a partial and noisy observational time series and a physics-based but imperfect parametric model are available, as in many realistic situations. The BAMCAFE combines ML with DA to improve the predictions utilizing the standard ensemble forecast by running the imperfect parametric model forward. Different from many purely data-driven methods, the BAMCAFE  takes advantage of the available imperfect parametric model to extract useful information that feeds into a ML model via DA\@.
The BAMCAFE starts with a Bayesian ensemble DA that aims to reduce the observational noise and recover the time series of the unobserved variables, which often allow the resulting assimilated trajectories to more accurately represent the underlying dynamics of nature than the imperfect parametric model. Therefore, if these assimilated trajectories are used to build a new forecast model, then the associated forecast is expected to be improved. Since it is in general very challenging to write down a set of physics-informed parametric equations to describe the time evolution of these assimilated trajectories,  ML models are used to characterize the underlying dynamics in the BAMCAFE framework.
Next, in addition to forecasting the optimal point-wise value, the BAMCAFE is also designed to quantify the forecast uncertainty. The forecast uncertainty in the BAMCAFE is represented by a non-Gaussian PDF, which is computed via an inexpensive algorithm from a mixture distribution. Such a non-Gaussian distribution is particularly appropriate for characterizing the uncertainty in complex turbulent systems in the presence of intermittency and extreme events. Then in the forecast initialization stage of the BAMCAFE approach, a generalized ensemble DA is utilized, which provides an ensemble of time series that serve as the input of the ML model. Subsequently, the ML forecast is carried out for each ensemble member to obtain both the point-wise forecast value and the forecast uncertainty. Overall, the BAMCAFE is computationally efficient in both the training and forecasting stages. It collects the useful information from the partial and noisy observations as well as the imperfect physics-informed model to facilitate the ML ensemble forecast. \\ 

The rest of the paper is organized as follows. The  BAMCAFE algorithm is developed in Section~\ref{sec: BAMCAFE}. Section~\ref{sec: numerics} includes nonlinear and non-Gaussian test examples comparing the forecast skill of the BAMCAFE algorithm and the associated imperfect models. The paper is concluded in Section~\ref{sec: discussion_conclusion}.

\section{The Bayesian Machine Learning Advanced Forecast Ensemble (BAMCAFE) Framework}\label{sec: BAMCAFE}
\subsection{Overview}

The ensemble forecast is a popular prediction approach for turbulent systems that employs a collection, known as the ``ensemble'', of multiple individual forecasts from a parametric model. A skillful ensemble forecast requires an accurate representation of the underlying dynamics as well as a reliable forecast initialization. DA is used to generate a more accurate initialization by combining partial and noisy observations with the given imperfect model. However, the forecast error can grow up quickly as time evolves due to the model bias, which often results in an inaccurate quantification of the forecast uncertainty as well.

\begin{figure*}[ht]
	\centering
	\includegraphics[width=\textwidth]{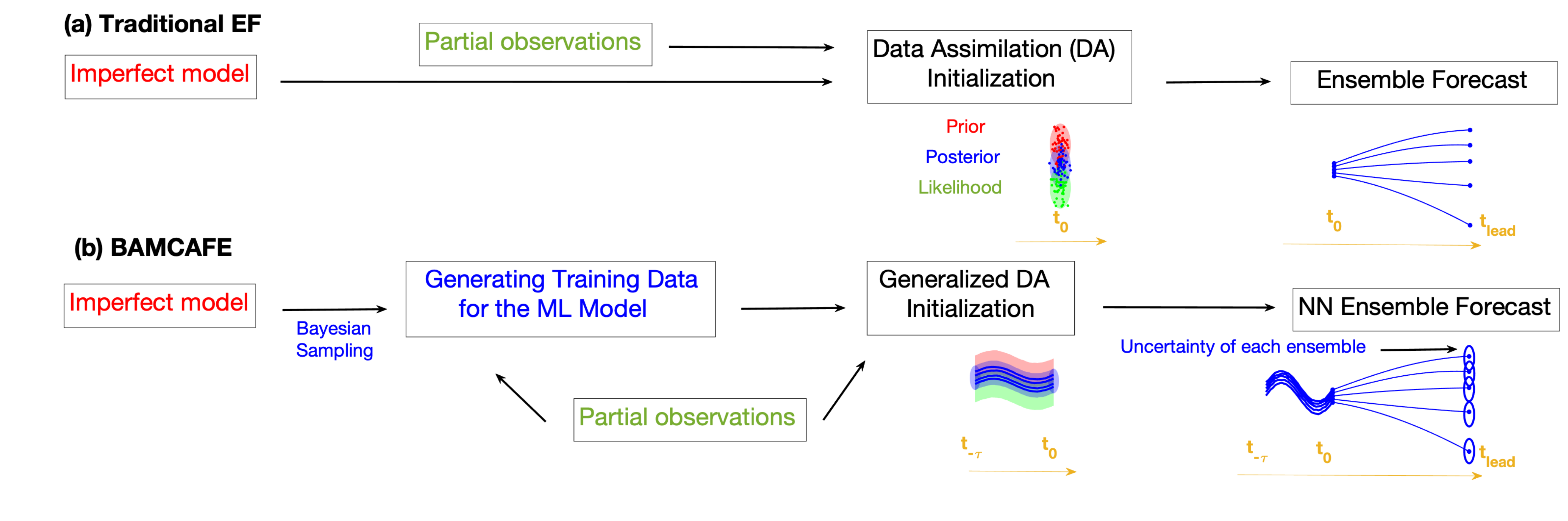}
	\caption{A schematic illustration of the traditional physics-informed parametric model based ensemble forecast and the BAMCAFE approach.}
	\label{fig: schematic_fig}
\end{figure*}

In the  \underline{Ba}yesian \underline{M}a\underline{c}hine \underline{L}earning \underline{A}dvanced \underline{F}orecast \underline{E}nsemble (BAMCAFE) framework, the intrinsic error in the imperfect physics-based forecast model is alleviated by training a ML model (e.g., a NN) based on a set of assimilated trajectories, which are obtained by applying a Bayesian sampling method (i.e., a Bayesian ensemble DA) to the imperfect physics-based  model with the help from the available partial and noisy observational time series. Combining the information from both the imperfect model and the noisy observations, the assimilated trajectories achieve trajectory-wise improvement compared with the signals generated from the imperfect model in terms of both the dynamical and statistics features. 
Specifically, the BAMCAFE involves the following four steps:
\begin{enumerate}
	\item Generating the ML training data using a Bayesian sampling approach. \label{step_1}
	\item Training a ML model (e.g., a NN) utilizing the training data from Step~\ref{step_1}.\label{step_2}
	\item Employing a generalized DA for the initialization of the ML model.
    \item Applying a ML ensemble forecast.
\end{enumerate}
The generalized DA in Step 3 aims at recovering a short piece of time series before the initial time instant for forecast, serving as the initialization of the ML model. It plays a similar role as the traditional physics-based parametric model forecast, but the initialization is not just at a single time instant.
Figure~\ref{fig: schematic_fig} includes a schematic illustration of the traditional physics-informed parametric model based ensemble forecast and the BAMCAFE approach. The details of each of the four steps in the BAMCAFE algorithm will be explained in the following subsections.

\subsection{Generating the ML training data using a Bayesian sampling approach}\label{sec: generating_training_data}

The first step of the BAMCAFE algorithm is to generate a set of trajectories that will be used to train the ML model. This is achieved by exploiting a Bayesian sampling approach that takes into account the information from both the given imperfect physics-based parametric model and the available partial and noisy observational time series.  Denote by $\x_j$ the state variable and $\y_j$ the observational vector at time $t_j$, where the dimension of $\y_j$ is often smaller than that of $\x_j$ due to the partial observations. The training data for the ML model is generated from the following conditional distribution,
\begin{equation}\label{eq: conditional_distribution}
    p(\x_{0}, \x_{1}, \dots, \x_N | \y_1, \y_2, \dots, \y_N), \qquad 0 \le j \le N,
\end{equation}
where~$\{\y_1, \y_2, \dots, \y_N\}$ stands for all the available observations in a given time interval. For the rest of this paper, we define~$\y_{i:j} = (\y_{i}^{\top}, \y_{i + 1}^{\top}, \dots \y_{j}^{\top})^{\top}$ for $0 \le i \le j$. In the Bayesian framework, the statistical information provided by the available imperfect parametric model is called the prior distribution while the conditional distribution~\eqref{eq: conditional_distribution} is named as the posterior distribution. Many existing Bayesian sampling algorithms can be applied to compute the posterior distribution~\eqref{eq: conditional_distribution}~\cite{tong2020mala, bedard2017hierarchical, johnson2013component, ottobre2016function, agapiou2017importance}. In this work, the ensemble Kalman smoother (EnKS)~\cite{evensen2000ensemble}, which is one of the commonly used DA approaches, is adopted to sample from the posterior distribution. The technique details of sampling from the posterior distribution utilizing the EnKS are summarized in the Appendix~\ref{sec: SI_EnKF}. Denote by~$\{\x_{0| N}^{k}, \ldots, \x_{N| N}^{k}\}$ for $k=1, \dots, K$ the resulting $K$ ensemble members from the EnKS, each of which is a time series. The trajectories of these ensemble members will be used as the training data for the ML model. By using the Bayesian DA, the model error is mitigated in these posterior time series due to the extra information from observations. In addition, the training data are now also available for the unobserved variables, without which it is quite challenging to apply the ML models to predict the entire system.

It is important to note that despite the ensemble mean time series from the EnKS $\{{\bmu}_{0 | N},\ldots, {\bmu}_{N | N}\}$ being a widely used surrogate for the true signal as in the standard reanalysis approaches, the ensemble mean time series is not always a suitable training data for the ML model because the fluctuation of the original turbulent dynamics is smoothed out in the ensemble mean time series. In other words, the dynamical and statistical features of the underlying dynamics are not fully reflected in the ensemble mean time series. Different from the smoother mean, each ensemble member~$\{\x^{k}_{0 | N},\ldots,\x^{k}_{N | N}\}$, referred as a sampled trajectory, is more appropriate for training the ML models. This is one of the fundamental differences between the training data in the BAMCAFE framework and the traditional reanalysis outcome that is often given by the ensemble mean time series. A comparison between the time series of the ensemble members and the smoother mean as well as the associated ML forecast results will be presented in Section~\ref{sec: sampling_smoothing}. 

It is also worthwhile to note that the EnKS is a more suitable method than the ensemble Kalman filter (EnKF), which takes into account only the information in the past, to create the ML training  time series.  The state estimation using the EnKF is often less accurate than that using the EnKS, especially in the presence of strong turbulence, intermittency and extreme events~\cite{evensen2000ensemble, chen2020efficient}. Since the training procedure is offline, where the observational time series in a given interval $\y_{1:N}$ is in hand, it is natural to adopt the EnKS for creating the training data for the ML models.

\subsection{Training a ML model}\label{sec: training_ML_models}

Given the sampled trajectories from Step~\ref{step_1}, the second step of the BAMCAFE algorithm is to build a new model that captures the key features of these sampled trajectories. Since it is often quite difficult to develop a physics-based parametric model that perfectly captures the underlying dynamics of these sampled trajectories, it is natural to employ a ML model  to characterize their time evolutions. For the convenience of discussion, a NN will be utilized as the ML model throughout the paper.
It is worthwhile to note that numerous works have been shown that NNs can well approximate many complex dynamics~\cite{chen1995universal, hornik1993some, sonoda2017neural}.

\begin{figure*}[ht]
	\centering
	\includegraphics[width=\textwidth]{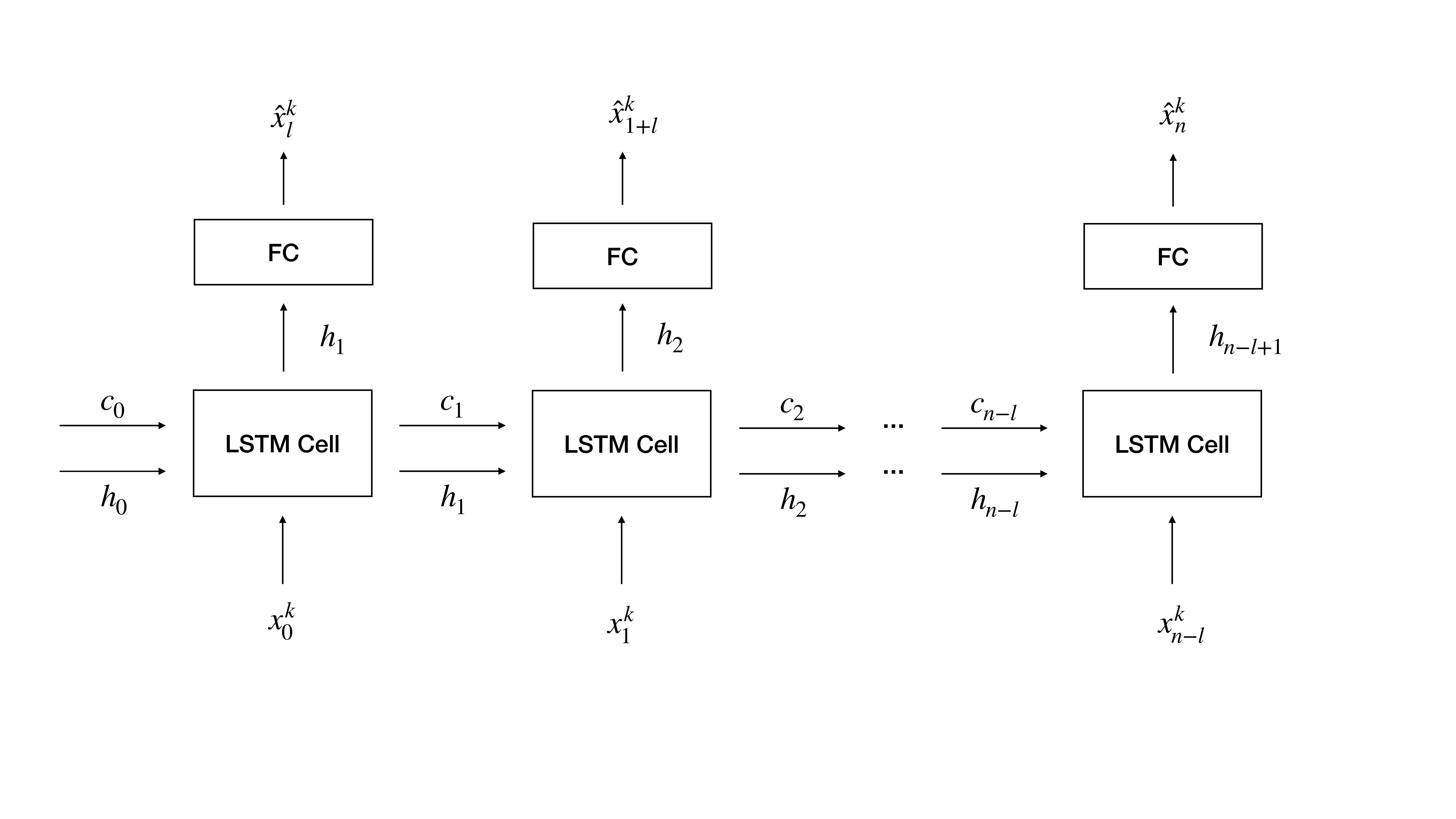}
	\caption{Structure of the one layer LSTM model used in this work. FC means a fully-connected layer.}
	\label{fig: LSTM_arc}
\end{figure*}

Denote by $\x^k_{0:N}$ the $k$-th member of the sampled trajectories from $t_0$ to $t_N$, as a simpler notation for $\{\x_{0| N}^{k}, \ldots, \x_{N| N}^{k}\}$. The entire sampled trajectory is further separated as the training and the validation periods, denoted by~$\x^k_{0:N_{\textrm{tr}}}$ and~$\x^k_{N_{\textrm{tr}}:N}$, respectively. The NN model is defined as $g(\x; \btheta)$, where $\btheta$ is a set of trainable parameters in the network. The ML task is to predict the value that is $l$ steps forward in time, which can be expressed as follows
\begin{equation}\label{eq: ML_model}
	\hat{\x}^k_{n + l} = g(\x^k_{0:n}; \btheta), ~\textrm{for } N_{0} \le n \le N_{\textrm{tr}} - l,
\end{equation}
where $l$ is the lead time or forecast horizon and $N_{0}$ is the length of the initial period of the NN\@. We aim to introduce the framework instead of sophisticated NN architectures, a one-layer LSTM model followed by a fully connected layer will be used in the numerical experiments of this work, the NN structure of which is shown in Figure~\ref{fig: LSTM_arc}. The LSTM here can be replaced by other non-parametric models such as convolutional neural networks depending on the applications. Note that the right-hand side of~\eqref{eq: ML_model} is written in a general form, in which the recurrent neural network can take varying length sequences as the input. A simpler version of the general form~\eqref{eq: ML_model} replaces $\x^k_{0:n}$ by $\x^k_{n-N_\textrm{init}:n}$ on the right-hand side, which is widely used in practice. The loss function used in the training phase is given as follows,
\begin{equation}\label{eq: loss_function}
	J(\btheta) = \sqrt{\frac{1 }{K(N_{\textrm{tr}} - l + 1 - N_{0}) } \sum_{k = 1}^{K} \sum_{n = N_{0}}^{N_{\textrm{tr}} - l} \|\hat{\x}^k_{n}(\btheta) - \x^k_{n} \|^2}.
\end{equation}
The loss function~\eqref{eq: loss_function} is the mean-squared error, which is a commonly used loss function in time series predictions. Again, other loss functions can be utilized in the BAMCAFE framework. Note that the error $\hat{\x}^k_{n}(\btheta) - \x^k_{n}$ evaluated on validation set is crucial in quantifying the uncertainty  in the forecast stage, which will be discussed in Section~\ref{sec: ML_Forecast}.

\subsection{Employing a generalized DA for the initialization of the ML model}
With the trained NN in hand, what remains is to apply the BAMCAFE for the ensemble forecast. The goal is to predict both the point-wise value and the associated uncertainty at the lead time~$l$, i.e., to predict~$\x_{N + l}$ given an observational sequence~$\y_{1:N}$.

Following~\eqref{eq: ML_model}, the predicted value for each ensemble member is given by
\begin{equation}\label{eq: predicted_value}
	\hat{\x}^k_{N + l} = g(\x^k_{N - L_{\textrm{init}}:N}; \btheta^*),
\end{equation}
for $k=1,\ldots, K_p$, where $K_p$ is the total number of the ensembles used in prediction that can be different from the number of the training trajectories $K$.
In~\eqref{eq: predicted_value}, $\btheta^*$ is the value minimizing the loss function obtained from the Step~\ref{step_2} while $L_{\textrm{init}}$ is the length of the initial period. Adopting a period of time series $\x^k_{N - L_{\textrm{init}}:N}$ as the input is necessary for the LSTM and other networks to take into account the memory effects of the dynamics that significantly facilitates the forecast, which is also consistent with the input data in~\eqref{eq: ML_model}.

Similar to the training data, only partial observations $\y_{1:N}$ are available for the state variable $\x_{1:N}$ at the forecast initialization stage. Therefore, an initialization procedure has to be applied to obtain $\x^k_{N - L_{\textrm{init}}:N}$ that is the prerequisite for implementing~\eqref{eq: predicted_value}. This initialization can be regarded as a generalization of the traditional DA because 1) the initial condition $\x^k_{N - L_{\textrm{init}}:N}$ can be obtained via a Bayesian ensemble DA formula; and 2) the value at the end point $\x^k_{N}$ is exactly the initial values used in the traditional ensemble forecast. The main difference is that the initial condition in the traditional approaches is a group of point-wise samples at only the time instant~$t_{N}$ yet those in the BAMCAFE framework is a group of time series  from $t_{N - L_{\textrm{init}}}$ to $t_N$. The generalized DA for the initialization of the NN is carried out utilizing the following Bayesian formula
\begin{equation}\label{eq: conditional_distribution_initilization}
    p(\x_{N - L_{\textrm{init}}}, \x_{N - L_{\textrm{init}}+1}, \dots, \x_N | \y_1, \y_2, \dots, \y_N), \qquad 0 \le j \le N,
\end{equation}
where the prior information is still provided by the imperfect parametric model, as in~\eqref{eq: conditional_distribution}. Again, the EnKS is applied in this work to sample the trajectories from the posterior distribution~\eqref{eq: conditional_distribution_initilization}, serving as the initialization of the NN forecast model in~\eqref{eq: predicted_value}. Note that the value of $\x^k_{N}$ from the EnKS is the same as that by applying the EnKF but those $\x^k_{N - L_{\textrm{init}}}, \x^k_{N - L_{\textrm{init}}+1}, \dots, \x^k_{N-1}$ are statistically more accurate utilizing the EnKS\@.

\subsection{Applying a ML ensemble forecast}\label{sec: ML_Forecast}
Given the initialization from the generalized DA, it is ready to run~\eqref{eq: predicted_value} for the ensemble forecast.

The ensemble mean, which is the most widely used surrogate for the point-wise prediction value, is calculated as follows,
\begin{equation}\label{eq: Ensemble_mean}
	\bar{\hat{\x}}_{N + l} = \frac{1}{K_p} \sum_{k=1}^{K_p} \hat{\x}^k_{N + l},
\end{equation}
where $\hat{\x}^k_{N + l}$  is calculated from~\eqref{eq: predicted_value}.
 Note that the mode or other statistical measures can also be used as the optimal point-wise prediction value depending on the quantity of interest of the applications.

In addition to the ensemble mean, predicting the uncertainty is also essential, especially for complex turbulent dynamics with intermittency and extreme events. Unlike many ML approaches that focus only on the optimal point-wise forecast value, the BAMCAFE also provides the quantification of the forecast uncertainty. The total forecast uncertainty contains two parts. The first part is the ensemble spread of the point-wise forecasts, namely the $\hat{\x}^k_{N + l}$ with $k=1,\ldots K_p$. The second and indispensable component is the intrinsic uncertainty associated with each ensemble member. The ensemble spread of the point-wise forecasts is the dominant source of the uncertainty at short lead times, where the spread comes from DA. As the lead time increases, the point-wise forecast becomes less accurate, the associated error of which contributes to the total uncertainty of the forecast. Because of this, the BAMCAFE exploits the validation error obtained in the ML training period as the measurement of the forecast uncertainty associated with each ensemble member. Such a simple criterion is a natural choice for quantifying the forecast uncertainty since it represents the residual part of the dynamics which cannot be well characterized and forecasted by the LSTM model. 
Some quantitative studies between the forecast uncertainty in the LSTM forecast based on the validation error and that in the ensemble forecast using the perfect parametric model will be illustrated in Section~\ref{sec: validation}.
Note that the validation error here can be replaced by the training error, assuming the NN is appropriate, i.e., without being overfitted or underfitted.
Specifically, the forecast uncertainty associated with the BAMCAFE is represented by a non-Gaussian PDF. This non-Gaussian PDF is constructed by a mixture distribution, where each mixture component is another non-Gaussian distribution that is associated with one forecast ensemble member in~\eqref{eq: predicted_value}. 
The $k$-th mixture component is given by adding the point-wise forecast value $\hat{\x}^k_{N + l}$ to a non-Gaussian distribution $\epsilon$,
\begin{equation}\label{eq: mean}
	p(\x_{N + l}^k) = \hat{\x}^k_{N + l} + \epsilon,
\end{equation}
where  $\epsilon$ is the distribution of the validation error,
\begin{equation}\label{eq: residual}
	\epsilon \sim \textrm{PDF of } \Big(\x^{k'}_{n + l} - g(\x^{k'}_{N_{\textrm{tr}}:n}; \btheta^*)\textrm{,~~ for } k' = 1, \dots, K,~~ N_{\textrm{tr}} + 1\le n \le N - l\Big).
\end{equation}

As a final remark, a connection can be built between using the parametric model and the BAMCAFE algorithm to characterize forecast uncertainty. In fact, the ensemble ML forecast algorithm of the BAMCAFE can be regarded as a surrogate of the ensemble forecast using physics-informed parametric models but using a different approach of computing the forecast uncertainty. To see such a connection, suppose that we have an ensemble of the forecast value at time $t_{N+l}$. Consider the following mean-fluctuation decomposition of the forecast value for each ensemble with an index $k$,
\begin{equation}\label{MF_Decomposition}
    \hat{\x}^k_{N + l} = \bar{\hat{\x}}_{N + l} + (\hat{\x}^k_{N + l} - \bar{\hat{\x}}_{N + l}),
\end{equation}
where the first term on the right-hand side is the ensemble mean and the second term is the fluctuation. The uncertainty is essentially given by the statistics, namely the PDF, of the fluctuation part. In the traditional forecast using physics-informed parametric models, each ensemble member is given directly by running either stochastic or deterministic chaotic models forward. The results of such an ensemble forecast can be directly used to form the forecast PDF. Nevertheless, in most of the ML forecasts, the ML model/mapping is deterministic and non-chaotic. The ML forecast output is typically a deterministic value provided by a certain averaging process (i.e., the optimal mapping) inside the complicated ML architecture. There is a residual from such an averaging process, which is characterized by the validation error. The validation error for each sample is essentially the second part on the right-hand side of~\eqref{MF_Decomposition}. In other words, the validation error mimics the ensemble spread in the traditional ensemble forecast using physics-informed parametric models. If the training and testing data have the same features, then it is expected that the residual in the testing period should be similar to that in the validation period. Therefore, it is natural to use the validation error as a characterization of the uncertainty. Notably, a more accurate ML model in terms of representing the perfect model dynamics is expected to provide a closer result of the forecast uncertainty as the perfect model. This fact implies that the Bayesian DA is crucial in creating an improved data set for training the ML model.

\section{Test Examples}\label{sec: numerics}

This section will illustrate how the BAMCAFE improves ensemble forecast results by employing testing models that mimic many desirable features of complex turbulent systems in reality. In each test case, we demonstrate a perfect model with specific parameters that produces the true dynamics. The baseline is the ensemble forecasts using one or two commonly used imperfect parametric models. Based on 50 sampled trajectories, a NN model, as was shown in Figure~\ref{fig: LSTM_arc}, with a one-layer LSTM network followed by a fully connected layer is trained. The `Adam' optimization algorithm and the mean-squared error loss function~\eqref{eq: loss_function} are utilized to train different LSTM models for different lead times. The details of hyperparameters are included in Appendix~\ref{sec: SI_Hyperparameters}.

The two test experiments used below will emphasize different aspects of the BAMCAFE framework. In the first case, the goal is to illustrate that the BAMCAFE algorithm can improve the forecasting skill compared to the typical reduced-order imperfect models with either bare truncation or stochastic parameterization. In the second case, it is highlighted that the BAMCAFE facilitates the quantification of the forecast uncertainty in a strongly turbulent system with intermittency and extreme events. 

\subsection{General experiment setup}
The true dynamics are integrated from the perfect model using the Euler-Maruyama scheme~\cite{gardiner2004handbook}, with the integrated time step~$\Delta t$ being 0.001 for both the perfect and imperfect models. The discrete observations are collected for every 0.05 time units.

The two criteria for quantifying the overall point-wise prediction skill are the root-mean-square error (RMSE) and the pattern correlation (Corr) between the prediction $\hat{x}$ and the true signal $x$. These criteria are defined as~\cite{hyndman2006another}:
\begin{equation}\label{eq: Skill_Scores}
\begin{split}
  \mbox{Corr} &= \frac{\sum_{i=1}^I(\hat{x}_{i}-\bar{\hat{x}})(x_i-\bar{x})}{\sqrt{\sum_{i=1}^I(\hat{x}_{i}-\bar{\hat{x}})^2}\sqrt{\sum_{i=1}^I(x_{i}-\bar{\hat{x}})^2}},\\
  \mbox{RMSE} &=  \sqrt{\frac{\sum_{i=1}^I(\hat{x}_{i}-x_{i})^2}{I}},
\end{split}
\end{equation}
where $\bar{\hat{x}}$ and $\bar{x}$ are the averages of the scalar quantities $\hat{x}_i$ and $x_i$ over $i=1,\ldots,I$, respectively.
In general, if RMSE is below one standard deviation of the true signal and Corr is above the threshold value Corr $= 0.5$, then the prediction is said to be skillful.

On the other hand, the relative entropy (also known as the Kullback–Leibler divergence) is one of the most suitable measurements that quantifies the accuracy of the forecast uncertainty~\cite{kleeman2011information, majda2002mathematical, xu2007measuring, branicki2013non}. It measures the statistical distance between the PDF $p(\x)$ associated with the perfect model ensemble forecast and that $p(\hat{\x})$ associated with either the imperfect parametric model ensemble forecast or that from the BAMCAFE at a given time instant. The relative entropy is defined as~\cite{kullback1951information, kullback1987letter, cover1991entropy}:
\begin{equation}\label{eq: Relative_Entropy}
\mathcal{P}(p(\x),p(\hat{\x})) = \int p(\x)\log\left(\frac{p(\x)}{p(\hat{\x})}\right).
\end{equation}
The relative entropy is zero if the two PDFs equal with each other. The relative entropy increases monotonically as the difference between the two PDFs becomes large.
\subsection{Improving the ensemble forecasts from the reduced-order models with  bare truncation or stochastic parameterization}
The first numerical test example is the two-layer Lorenz 96 model, which aims at showing that the BAMCAFE algorithm can improve the forecasting skill compared to the typical reduced order imperfect models with either bare truncation or stochastic parameterization~\cite{majda2009mathematical,grooms2014stochastic,grooms2014stochastic2,majda2014new}.

\subsubsection{The perfect model}
\begin{figure*}[ht]
	\centering
	\includegraphics[width=\textwidth]{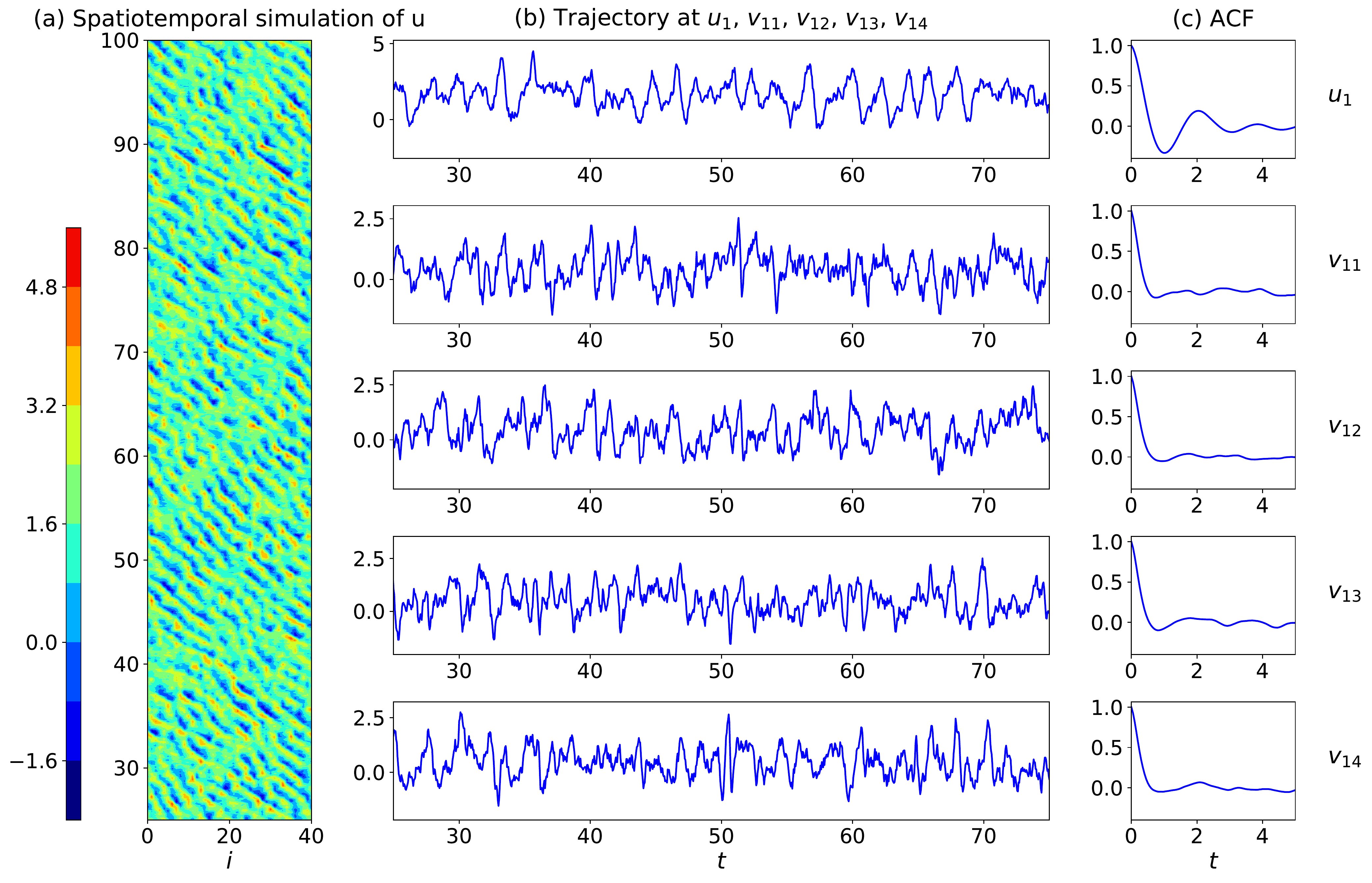}
	\caption{The two-layer L96 model~\eqref{eq: L96_model} with parameters in~\eqref{eq: L96_model_para}. Panel (a): The spatiotemporal patterns of large-scale variables $u_i$. Panel (b): the time series of the large-scale variable $u_1$ and the associated small-scale variables $v_{11},\ldots,v_{14}$. Panel (c): the ACFs corresponding to the variables in Panel (b).}
	\label{fig: L96_true}
\end{figure*}

The two-layer Lorenz 96 (L96) model is a conceptual representation of geophysical turbulence that is widely used in numerical weather forecasting as a testbed for DA and parameterization~\cite{lorenz1996predictability,lee2017multiscale,wilks2005effects,arnold2013stochastic}. The model mimics a coarse discretization of atmospheric flow on a latitude circle, which exhibits complex wave-like and chaotic behavior. It illustrates the interactions between small-scale fluctuations and larger-scale motions schematically. The noisy version of the model reads

\begin{subequations}\label{eq: L96_model}
	\begin{align}
		\frac{d u_{i}}{d t} &= \left(- u_{i - 1}\left(u_{i - 2}-u_{i + 1}\right)-u_{i} + f - \frac{h c}{J} \sum^{J}_{j=1} v_{i, j} \right) + \sigma_{u_i} \dot W_{u_i}, \quad i = 1, \dots, I, \\
		\frac{d v_{i, j}}{d t} &= \left(-bc v_{i, j + 1}\left(v_{i, j + 2} - v_{i, j - 1}\right) - cv_{i, j} + \frac{hc}{J} u_{i} \right) + \sigma_{v_{i, j}} \dot W_{v_{i, j}}, \quad j=1, \dots, J,
\end{align}
\end{subequations}

where $I$, $J$, $f$, $h$, $c$, $b$, $\sigma_{u_i}$ and $\sigma_{v_{i,j}}$ are given parameters while $\dot W_{u_i}$ and $\dot W_{v_{i, j}}$ are white noise. The large-scale variables~$u_i$ are periodic in $i$ with $u_{i + I} = u_{i - I} = u_{i}$. The corresponding small-scale variables~$v_{i, j}$ are periodic in $i$ with $v_{i + I, j} = v_{i - I, j} = v_{i, j}$ and satisfy the following conditions in $j$: $v_{i, j + J} = v_{i + 1, j}$, and $v_{i, j - J} = v_{i - 1, j}$. The model discussed here uses variables $u_i$ to describe large-scale or slow movements which are resolved; small scales or rapid fluctuations represented by $v_{i,j}$ are often unresolved ones. The coupling of fast and slow variables is regulated by the parameter $h$. The parameter $c$ specifies how quickly the fast variables are damped in comparison to the slow variables. The parameter $f$ controls the magnitude of external large-scale forcing, while $b$ determines the amplitude of nonlinear interactions between the fast variables. As in the standard L96 model, we take $I=40$. There are $J = 4$ small-scale variables associated with each $u_i$. Thus, the total number of the state variables is 200. The constant forcing $f = 4$ makes the system to be chaotic. The parameters $h$, $c$, and $b$ are chosen in such a way that the small-scale variables have a comparatively significant impact on the large-scale ones. In other words, the perfect model only has a weak scale separation. The reason that we consider such a weak scale separation is that it better mimics the real atmosphere with chaotic/turbulent behavior,  the associated forecast of which is often very difficult. Finally, additional stochastic noise is added to the system, representing the contribution of the variables that are not explicitly modeled. The noise also interacts with the deterministic part via nonlinear terms, introducing additional complexity that mimics nature. To summarize, the parameters used in the perfect model~\eqref{eq: L96_model} are as follows,
\begin{equation}\label{eq: L96_model_para}
	I = 40, \quad J = 4, \quad h = 2, \quad c = 2, \quad b = 2, \quad  f = 4, \quad \sigma_{u_i} = \sigma_u = 1, \quad \sigma_{v_{i, j}} =  \sigma_v = 1.
\end{equation}
With these parameters, the spatiotemporal patterns are shown in Figure~\ref{fig: L96_true},  together with the time series of the variables at a fixed location $i=1$, i.e., the large-scale variable  $u_1$ and the small-scale variables $v_{1,1},\ldots, v_{1,4}$,  as well as the associated equilibrium PDFs and the autocorrelation functions (ACFs). The ACFs, which measure the memory of the system of each component, validate the weak scale separation adopted here. The average value of the ACF decay time, i.e., the decorrelation time, for large scales is 0.76.

\subsubsection{The imperfect parametric models}
Since the perfect knowledge of nature is never known or it is too complicated to be used in practice, approximate models with reduced dimensions or simpler structures are often utilized as the forecast models. Two imperfect (i.e., approximate) parametric models are introduced here, which are developed by applying two widely used approximation approaches as in many applications. The goal here is to make use of these imperfect models to predict the large-scale variables $u_{i}$ at all grid points, i.e., for $i=1,\ldots,I$.

The first imperfect model is the one-layer L96 (L96-1LYR) model, which is a reduced order model of the perfect system. It is also unknown as the bare truncation model, which is one of the simplest approximations by completely ignoring the small-scale variables. The L96-1LYR model reads
\begin{equation}\label{eq: L96_one_layer}
	\frac{d u_{i}}{d t} = \left(- u_{i - 1}\left(u_{i - 2}-u_{i + 1}\right)-u_{i} + f  \right) + \sigma_{u_i} \dot W_{u_i}, \quad i = 1. \dots, I.
\end{equation}
Bare truncation models are widely used in practice due to its reduced computational cost. In fact, the dimension of the L96-1LYR model~\eqref{eq: L96_one_layer} is only $40$, which is 5x smaller than that of the $200$-dimensional perfect model. However, the feedback from the small-scale variables $v_{i, j}$ to the large-scale variables $u_{i}$ is non-negligible here as in many real-world applications, which is the main source of the model error. Note that the bare truncation models may sometimes suffer from a finite-time blowup issue~\cite{majda2012fundamental}, which is however not a problem here as the model only contains advection and dissipation terms beyond a constant forcing.

The second imperfect parametric model is developed as follows. Instead of completely ignoring the small-scale variables, the equations of these unresolved scale variables are replaced by simple stochastic parameterized equations. The specific parameterization form adopted here follows the one that is widely-used in the stochastic parametrized extended Kalman filters (SPEKFs), which have been shown to be skillful for improving the DA and prediction skill~\cite{gershgorin2010improving,gershgorin2010test,majda2012filtering}. The stochastic parameterized imperfect model (L96-SP) has the following form:

\begin{subequations}\label{eq: L96_SP}
	\begin{align}
	\frac{d u_{i}}{d t} &= \left(- u_{i - 1}\left(u_{i - 2}-u_{i + 1}\right)-u_{i} + f - \frac{h c}{J} \sum^{J}_{j=1} v_{i, j} \right) + \sigma_{u_i} \dot W_{u_i}, \quad i = 1, \dots, I, \\
	\frac{d v_{i, j}}{d t} &= - \hat{d}_{i, j}(v_{i, j} - \hat{v}_{i, j}) + \hat{\sigma}_{v_{i, j}} \dot W_{v_{i, j}}, \quad j=1, \dots, J.~\label{eq: L96_SP_2}
	\end{align}
\end{subequations}

\begin{figure}[ht]
	\centering
	\includegraphics[width=\textwidth]{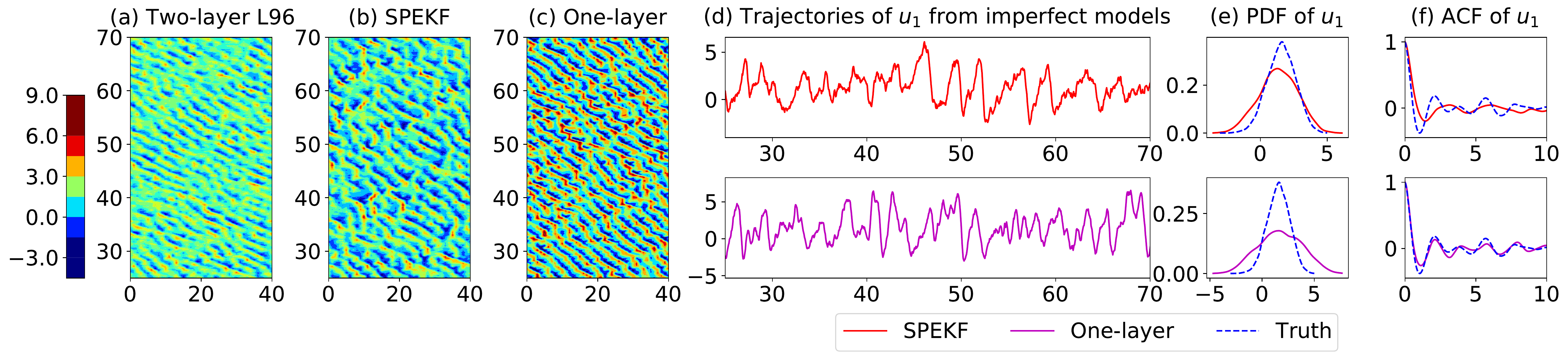}
	\caption{Comparison between the perfect model and two imperfect models associated with the two-layer L96 model. Panel (a)--(c): spatiotemporal patterns of the large-scale variables $u_i$. Panel (d)--(f): the time series, PDFs and ACFs of $u_1$, where the red curve shows those of the L96-SP~\eqref{eq: L96_SP}, the magenta curve shows those of the L96-1LYR~\eqref{eq: L96_one_layer} and the blue dashed curve corresponds to those of the perfect model~\eqref{eq: L96_model} as a reference.}
	\label{fig: L96_true_vs_wrong}
\end{figure}

In~\eqref{eq: L96_SP}, the unresolved small scales $v_{i, j}$ have been reduced to linear processes with only Gaussian additive noise providing statistically accurate feedbacks from the unresolved scales to the resolved ones. The parameters  in~\eqref{eq: L96_SP_2} can be calibrated by matching the mean, the variance, and the decorrelation time of $v_{i, j}$ with those in the perfect system, which provides the optimal Gaussian fit of each $v_{i, j}$ in~\eqref{eq: L96_SP_2} with that in the perfect model~\eqref{eq: L96_model}. The comparison between the perfect model~\eqref{eq: L96_model} and two imperfect models~\eqref{eq: L96_one_layer} and~\eqref{eq: L96_SP} is shown in Figure~\ref{fig: L96_true_vs_wrong}.
It is clear that both the imperfect parametric models capture certain features of the perfect dynamics but the model errors are also obvious. The L96-SP model is more accurate than the L96-1LYR model, as is expected. Note that the random numbers generated from the noise sources $\dot{W}_{u_i}$ and $\dot{W}_{v_{i,j}}$ when performing the ensemble forecast using both the perfect and imperfect models are different from the truth, which is a realistic setup.

\subsubsection{The experiment setup}
In this experiment, the two imperfect models (i.e., the L96-1LYR model and the L96-SP model) discussed above will be utilized to provide the baseline of the ensemble forecast results. They will then be utilized as the prior models to create sampled trajectories for training two LSTM models within the BAMCAFE framework. We refer to the LSTM models trained from the sampled trajectories generated from the L96-1LYR model~\eqref{eq: L96_one_layer} and the one from the L96-SP model~\eqref{eq: L96_SP} as the LSTM-L96-1LYR and the LSTM-L96-SP, respectively.

The observations are adopted only for the large-scale variables and are only on the even grid points. In other words, $u_2, u_4, \dots, u_{40}$ are the observed variables while $u_1, u_3, \dots, u_{39}$ and all the small-scale variables have no observations. The observation noise is assumed to be Gaussian with zero mean and a standard deviation of 1. The total length of the observational sequence is 8000 (400 time units) and is separated into the training and the validation data sets with a ratio 9:1. The testing set is the next 1600 observation steps (from 400 to 480 time units). The LSTM models are trained and are applied for forecast for all the large-scale variables. Sensitivity tests have been performed, which justifies that 400 units are sufficiently long to train the current one-layer LSTM models.

For the conciseness of presentation, the phrase ``imperfect models'' below always stands for the imperfect parametric models, i.e., the L96-1LYR model~\eqref{eq: L96_one_layer} and the L96-SP model~\eqref{eq: L96_SP}. The NN models will always have a prefix `LSTM'.

\subsubsection{The prediction skill}\label{sec: validation}
\begin{figure}[ht]
	\centering
	\includegraphics[width=\textwidth]{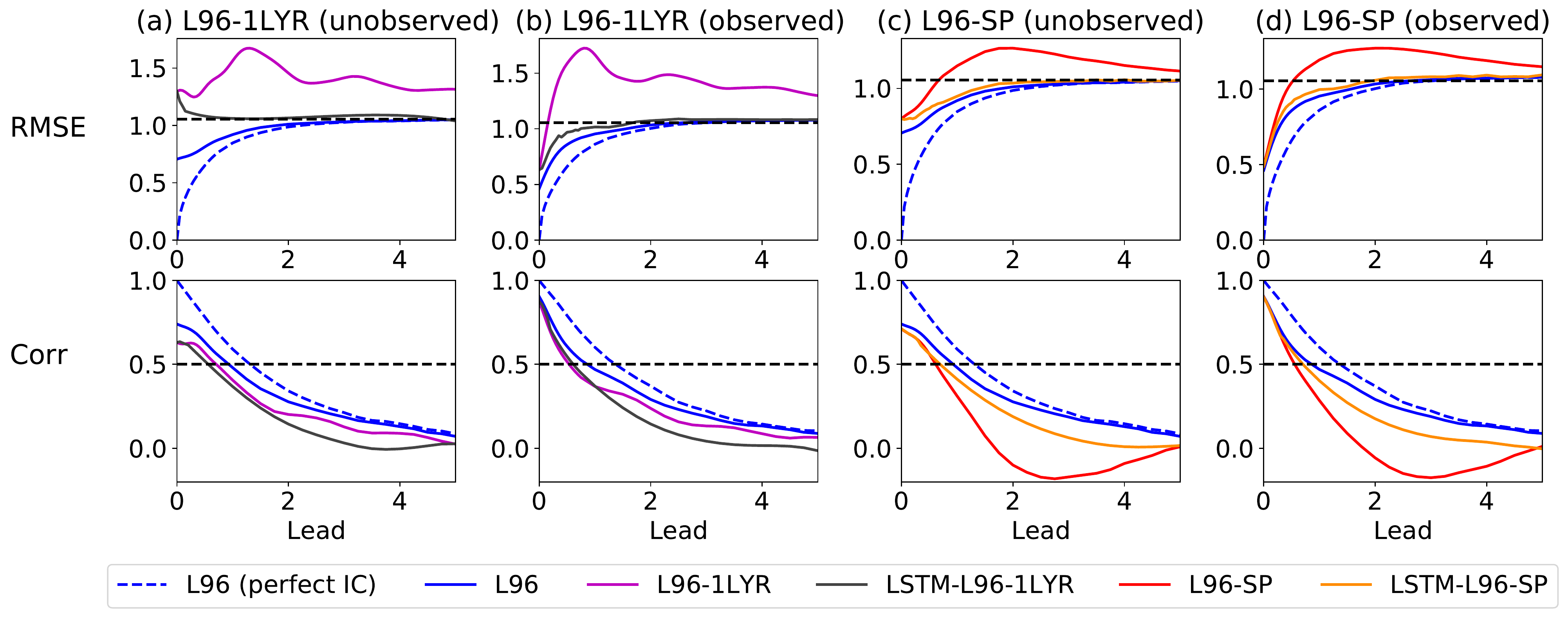}
	\caption{The RMSE and the Corr of the ensemble mean prediction as a function of lead time, where the true signal is generated from the two-layer L96 system~\eqref{eq: L96_model}. Panels (a)--(b): the skill scores averaged over all the 20 unobserved large-scale variables and those averaged over all the 20 observed variables using the one-layer L96 imperfect model~\eqref{eq: L96_one_layer} and the associated LSTM model. Panels (c)--(d): similar results but for the stochastic parameterized L96 model~\eqref{eq: L96_SP}. The blue dashed curves indicate the ensemble prediction using perfect model from perfect initial conditions as a reference. The black dashed lines in the RMSE panels represent the one standard deviation of the true signal and those in the Corr panels show the Corr = 0.5 threshold. }
	\label{fig: L96_RMSE_Corr}
\end{figure}

The RMSE and the Corr of the predicted ensemble mean time series related to the truth as a function of the forecast lead time are shown in Figure~\ref{fig: L96_RMSE_Corr}. The skill scores averaged over all the unobserved large-scale variables (in panel (a)) and over all the observed variables (in panel (b)) are shown using the perfect L96 model, the imperfect L96-1LYR model, and the LSTM-L96-1LYR model. Comparison between the perfect L96 model, the imperfect L96-SP model, and the LSTM-L96-SP model is shown in panel (c)--(d). In each panel, prediction from the perfect model with perfect initial condition (the true values initially at each time) is shown as a reference. When the two different imperfect models are compared to their corresponding LSTM models, almost all the skill scores from the LSTM models outperform those from the imperfect models, owing to the model error being significantly reduced in the training data. It is important to note that the skill scores of the LSTM models are also quite close to those associated with the perfect model, where the useful prediction lasts for more than 1 unit that is slightly longer than the average decorrelation time. In contrast, the errors in the imperfect models increase quickly. The forecast of the observational variables are more skillful than those in the unobserved large-scale variables at short lead times because of the larger initialization uncertainty of the latter. Besides, since the L96-1LYR model~\eqref{eq: L96_one_layer} completely ignores the damping effects of the small scales, the predicted values have larger amplitudes than the true values and the predictions have phase shifting tendencies.

\begin{figure}[ht]
	\centering
	\includegraphics[width=\textwidth]{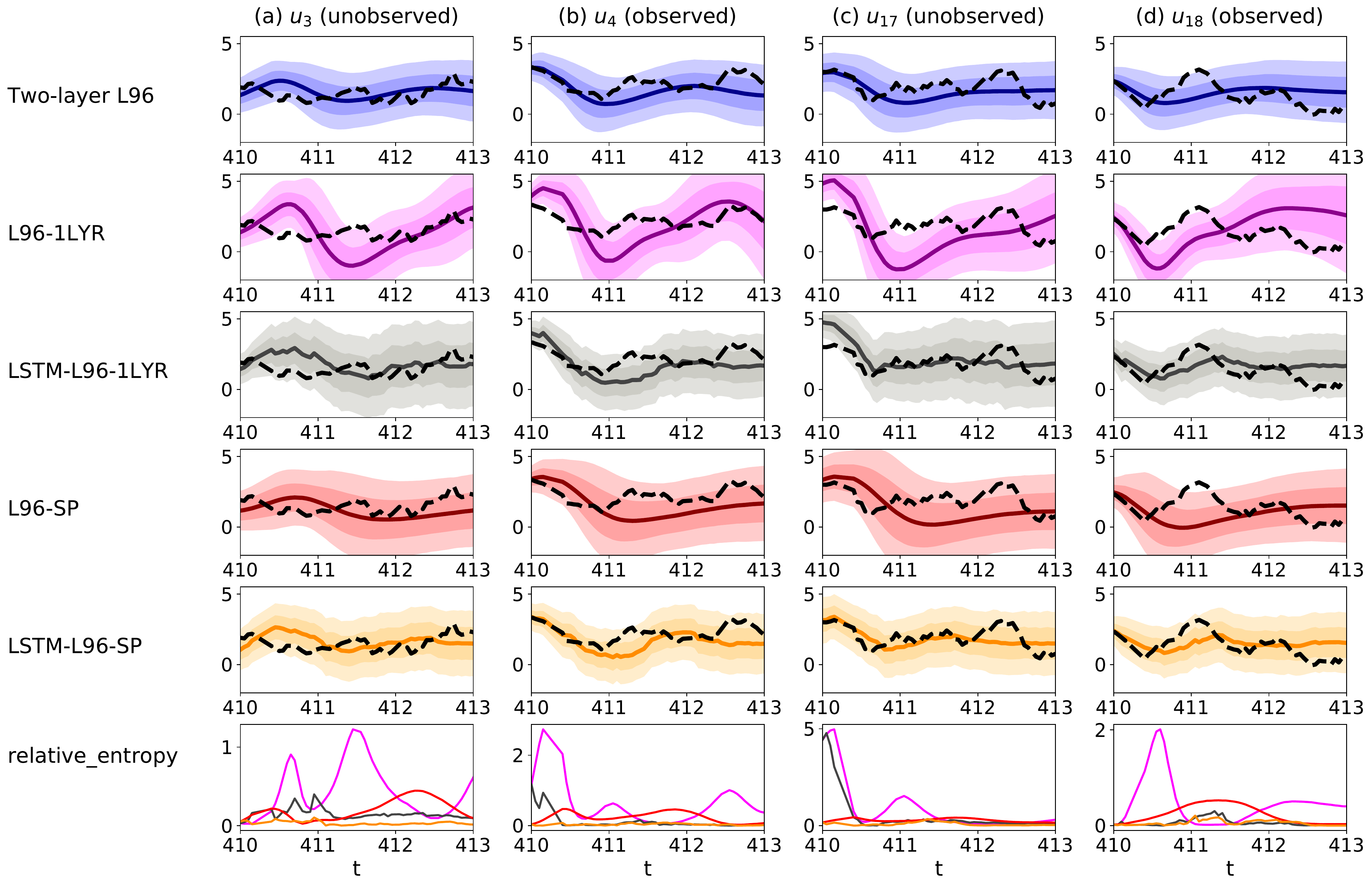}
	\caption{Ensemble forecasts of four large-scale variables of the two-layer L96 model. Columns (a) and (c) show the forecasts of the two unobserved variables $u_3$ and $u_{17}$ while Columns (b) and (d) show those of the two observed variables $u_4$ and $u_{18}$. Different rows (except the last one) show the forecasts using different models with assimilated initial conditions starting from $t=410$. The solid curves are ensemble mean for different models while the black dashed curve are the true trajectories. The dark and light shading areas show the one and two standard deviations of the uncertainty in the prediction. The last row show the time evolutions of the relative entropy between the perfect model ensemble forecast and the ensemble forecast using different imperfect parametric or LSTM models. The pink, gray, red, and orange curves show the results using the L96-1LYR model, the LSTM-L96-1LYR model, the L96-SP model and the LSTM-L96-SP model, respectively. }
	\label{fig: L96_ensemble_spread}
\end{figure}

Figure~\ref{fig: L96_ensemble_spread} depicts a case study in which the prediction begins at $t=410$ with the assimilate initial condition. The ensemble prediction is run for three time units forward in each subfigure. The time evolution of the forecast spread illustrated here is constructed using 2500 ensemble members for all physics-based models. For the LSTM models, the ensemble is made use of 50 components where each component is constructed by generating 50 samples from the distribution in~\eqref{eq: residual}. The black solid line is the ensemble mean and the dark and light shading areas show the one and two standard deviations of the uncertainty in the prediction, respectively.
First, it justifies the results in Figure~\ref{fig: L96_RMSE_Corr} that, compared to the imperfect models, the associated LSTM models can predict at longer lead times in terms of the ensemble mean forecast. For example, in Rows 2 and 3 of Panel (b), with the same initial condition, the ensemble mean prediction from the imperfect L96-1LYR model starts away from the truth quickly and ends up with larger uncertainty at~$t=413$. In contrast, the LSTM-L96-1LYR model can provide longer lead times for the ensemble mean prediction (nearly identical to the perfect model) and smaller uncertainty completely covering the true signal. Another example to illustrate this point is to compare Rows 4 and 5 of Panel (d). The LSTM-L96-SP model can predict the ascending trends from $t=410.5$ to $t = 411$ as the perfect model while the imperfect L96-SP model decays immediately.
Second, despite the possible initialization errors in the unobserved variables, the ensemble evolutions of the LSTM models can quickly adjust to follow those of the perfect model as is shown in Panel (c) row of the LSTM-L96-1LYR model, which is not the case for the associated imperfect L96-1LYR model.
Third, although the ensemble mean predictions from the two LSTM models are similar, which can also be found in score curves in Figure~\ref{fig: L96_RMSE_Corr}, the prediction uncertainty for the unobserved variables in LSTM-L96-1LYR is slightly larger due to larger variance of the training data generated by the bare truncation model which totally ignores the small scales. Nevertheless, the forecast uncertainty in the LSTM-L96-1LYR has been significantly reduced compared with the associated imperfect L96-1LYR model, indicating the success of the BAMCAFE in characterizing the forecast uncertainty even in the presence of a heavily biased imperfect parametric forecast model. The relative entropy skill scores (the last row) that describe the time evolutions of the forecast uncertainty of the imperfect and LSTM models related to the perfect model forecast also confirm the improvement of the LSTM models in quantifying the forecast uncertainty.

\begin{figure}[ht]
	\centering
	\includegraphics[width=0.7\textwidth]{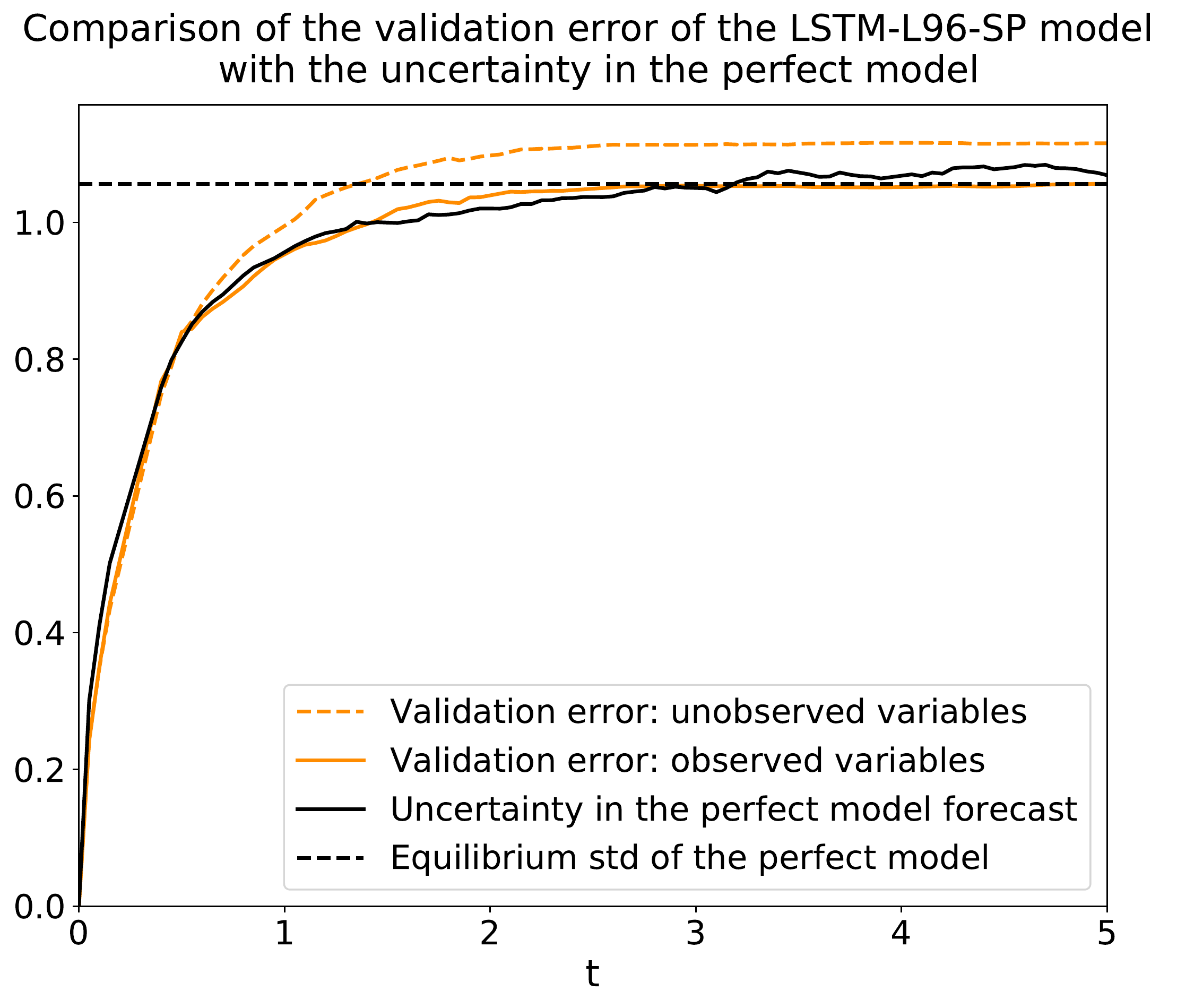}
	\caption{The time evolution of the validation errors using the LSTM-L96-SP model. The dashed and solid orange curves show the averaged validation error over the unobserved and observed large-scale variables, respectively. For comparison, the averaged forecast ensemble spread using the perfect model \eqref{eq: L96_model} is shown by the solid black curve. The `average' here means averaging the forecast deviation from the truth based on multiple forecasts starting from different time instants in a very long trajectory and averaging over all the $u_i$. Perfect initial conditions are used in the perfect model ensemble forecast to exclude the uncertainty due to the initial ensemble spread. The one standard deviation is used here to denote the averaged ensemble spread. The black dashed line is the one standard deviation of the forecast uncertainty at the statistical equilibrium state of the perfect model.}
	\label{fig: L96_training_error}
\end{figure}

The time evolution of the validation errors using the LSTM-L96-SP model are illustrated in Figure~\ref{fig: L96_training_error}, where the dashed and solid orange curves show the averaged validation error over the unobserved and observed large-scale variables, respectively. Recall in Section \ref{sec: ML_Forecast} that the validation error is the forecast uncertainty of each ensemble member in the LSTM model. For comparison, the averaged forecast ensemble spread using the perfect model \eqref{eq: L96_model} is shown by the solid black curve. The `average' here means averaging the forecast deviation from the truth based on multiple forecasts starting from different time instants in a very long trajectory and averaging over all the $u_i$. Perfect initial conditions are used in the perfect model ensemble forecast to exclude the uncertainty due to the initial ensemble spread. In such a way, both the LSTM-L96-SP and the perfect model illustrate the time evolution of the intrinsic uncertainty instead of the uncertainty resulting from the initializations.
It is shown that the time evolutions of the validation errors of the LSTM-L96-SP model are very similar to the averaged forecast uncertainty using the perfect model. Specifically, the validation errors of the LSTM-L96-SP model increase as a function of time and converge to the one standard deviation of the true signal at the statistical equilibrium state.  This means that the LSTM-L96-SP model succeeds in reproducing the time evolution of the forecast uncertainty of the perfect model.
The validation error in the unobserved large-scale variables is slightly larger than that in the observed ones. This is because the training data (i.e., the sampled trajectories) of the former contain slightly larger biases. Finally, despite the fact that Figure \ref{fig: L96_training_error} only compares the validation errors in the LSTM model with the perfect model using the perfect initial conditions, the LSTM-L96-SP also has almost the identical uncertainty spread as the perfect model forecast when DA is applied to the initialization, which can be seen from the second last row of Figure~\ref{fig: L96_ensemble_spread}.

\clearpage
\subsection{Quantifying the forecast uncertainty in a conceptual nonlinear model with strongly non-Gaussian features}
The purpose of this subsection is to demonstrate the skill of the BAMCAFE algorithm in quantifying the forecast uncertainty in a strongly turbulent systems with intermittency and extreme events.

\subsubsection{The perfect model}
The perfect model used here is the a nonlinear triad model with energy-conserving nonlinear interaction~\cite{majda2015statistical, chen2016filtering},
\begin{subequations}\label{eq: Triad_model}
\begin{align}
  \frac{du_1}{dt} &= \big(-\gamma_1 u_1 + L_{12} u_2 + L_{13} u_3 + I u_1 u_2 + F \big)  + \sigma_1 \dot W_1,\label{eq: Triad_u1}\\
  \frac{du_2}{dt} &= \big(- L_{12} u_1 - \frac{\gamma_2}{\delta} u_2 + L_{23} u_3 - I u_1^2 \big)  + \frac{\sigma_2}{\delta^{1/2}} \dot W_2,\label{eq: Triad_u2}\\
  \frac{du_3}{dt} &= \big(- L_{13} u_1 - L_{23} u_2 - \frac{\gamma_3}{\delta} u_3  \big) + \frac{\sigma_3}{\delta^{1/2}} \dot W_3,\label{eq: Triad_u3}
\end{align}
\end{subequations}
where $\gamma_i$, $L_{12}$, $L_{13}$, $L_{23}$, $I$, $F$, $\sigma_i$, and $\delta$ are given constants while $\dot W_i$ are independent white noise. This nonlinear triad system is a simple prototype nonlinear stochastic model that mimics structural features of low-frequency variability of general circulation models with non-Gaussian features~\cite{majda2009normal} and it was used to test the skill for reduced nonlinear stochastic models for fluctuation dissipation theorem~\cite{majda2010low}. The triad model~\eqref{eq: Triad_model} involves a quadratic nonlinear interaction between~$u_1$ and~$u_2$ with energy-conserving property that induces intermittent instabilities. On the other hand, the coupling between~$u_2$ and~$u_3$ is linear and is through the skew-symmetric term with coefficient~$- L_{23}$, which represents an oscillation structure of $u_2$ and $u_3$.

\begin{figure}[ht]
	\centering
	\includegraphics[width=\textwidth]{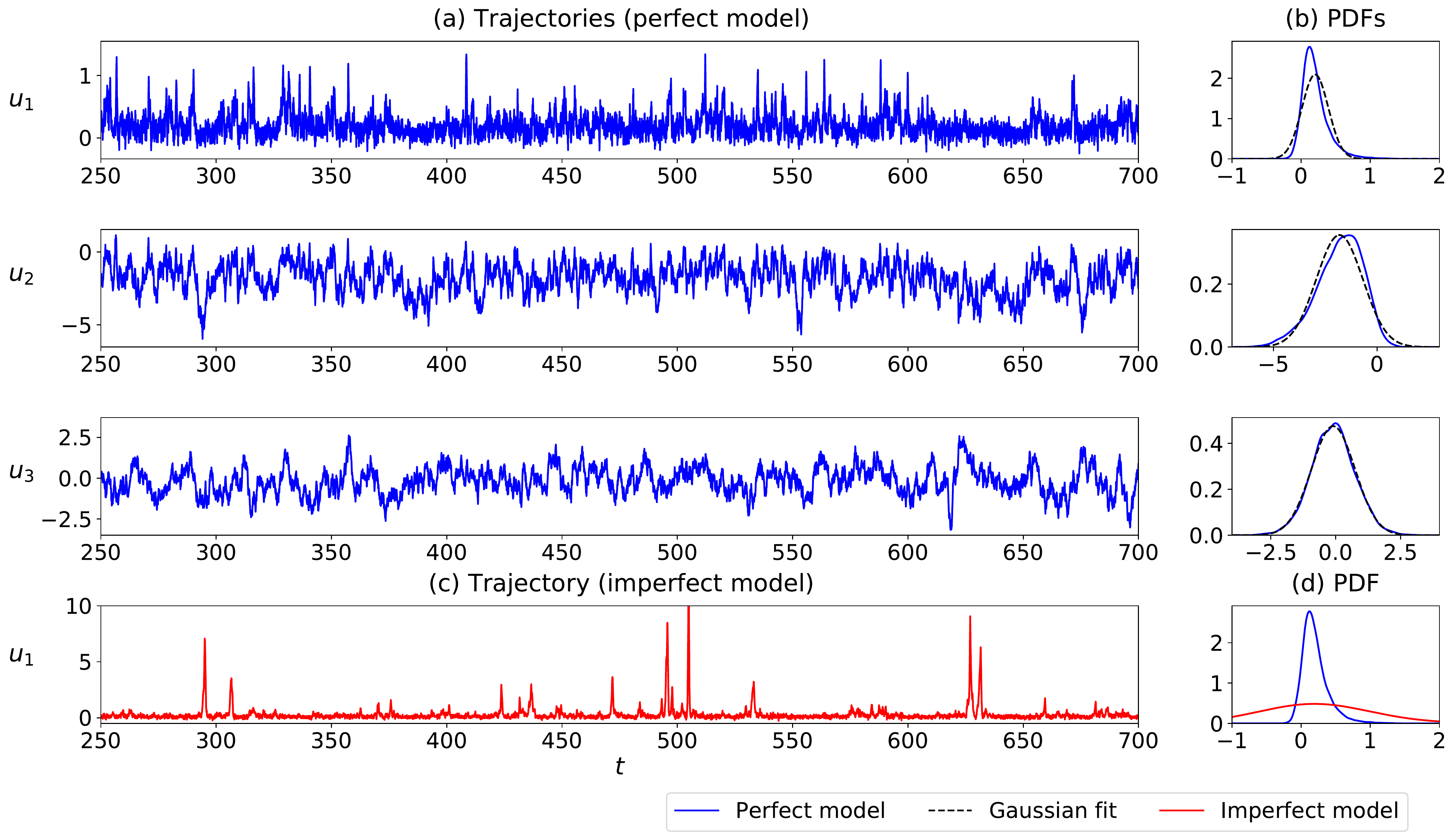}
	\caption{Panel (a): A realization of the triad model~\eqref{eq: Triad_model} with the parameters in~\eqref{eq: Triad_model_para}. Panel (b): the associated equilibrium PDFs, where the black dashed curves are the Gaussian fits of the true PDFs. Panel (c)--(d): A realization and the associated PDF of $u_1$ from the imperfect model~\eqref{eq: Triad_model_imperfect}.}
	\label{fig: Triad_trajectory}
\end{figure}

The parameter $\delta$ controls the scale separation of the system. If $\delta\ll 1$, then \eqref{eq: Triad_model} becomes a slow-fast system. If $\delta$ is of order $O(1)$, then all the three variables lie in a similar time scale.
This model has been widely used as a testbed for the DA and forecast of complex turbulent systems~\cite{majda2018model, chen2018efficient, chen2016filtering}. The following parameters are adopted for the nonlinear triad model~\eqref{eq: Triad_model_para} in this experiment,
\begin{equation}\label{eq: Triad_model_para}
	\begin{aligned}
	\gamma_1 = 2, \quad \gamma_2 = 0.2, & \quad \gamma_3 = 0.4, \quad L_{12} = 0.2, \quad L_{13} = 0.1  \quad L_{23} = 0, \quad I = 5, \quad \delta = 1 \\
	& \sigma_1 = 0.5, \quad \sigma_2 = 1.2, \quad \sigma_3 = 0.8, \quad F = 2
	\end{aligned}
\end{equation}
The blue curves in Figure~\ref{fig: Triad_trajectory} show one realization of the model where the model trajectories are given in Panel (a) and the equilibrium PDFs are given in Panel (b). The black dashed curves in Panel (b) are the Gaussian fits of the true equilibrium distributions. It is clear that strong intermittency and extreme events in the time series of $u_1$ lead to a highly non-Gaussian PDF with an one-sided fat tail. The strong nonlinear feedback from $u_1$ to $u_2$ also introduces a skewed PDF of $u_2$.

\subsubsection{The imperfect model}
In various applications~\cite{majda2009normal, majda2010low, chen2018efficient} , it is often assumed that $u_1$ is the observed variables while $u_2$ and $u_3$ represent the  unresolved processes. Therefore, approximate models are developed, which includes simple parameterizations for these two unresolved processes. One natural way of proposing an imperfect approximate model is to use linear stochastic processes to replace the nonlinear dynamics of $u_2$ and $u_3$~\cite{majda2012filtering}. The resulting model reads,
\begin{subequations}\label{eq: Triad_model_imperfect}
	\begin{align}
		\frac{du_1}{dt} &= \big(-\gamma_1 u_1 + L_{12} u_2 + L_{13} u_3 + I u_1 u_2 + F \big)  + \sigma_1 \dot W_1,\label{eq: Triad_imperfect_u1}\\
	  \frac{du_2}{dt} &= -d_{u_2}^M (v - \bar{u}_2^M) + \sigma_{u_2}^M \dot W_{u_2},\label{eq: Triad_imperfect_u2} \\
	  \frac{du_3}{dt} &= -d_{u_3}^M (v - \bar{u}_3^M) + \sigma_{u_3}^M \dot W_{u_3},\label{eq: Triad_imperfect_u3}
	\end{align}
\end{subequations}
Assume that the parameters in \eqref{eq: Triad_imperfect_u1} for the observed variable $u_1$ are the same as those in the perfect model~\eqref{eq: Triad_model} while the parameters in \eqref{eq: Triad_imperfect_u2}--\eqref{eq: Triad_imperfect_u3} for the unobserved variables $u_2$ and $u_3$ are calibrated by matching the mean, variance, and decorrelation time with those in~\eqref{eq: Triad_model}. Note that due to violation of the energy-conserving constraint of the nonlinear terms in \eqref{eq: Triad_model_imperfect},  the amplitude of $u_1$ from the imperfect model is much larger than the one from the perfect model. See Panel (c) in Figure~\ref{fig: Triad_trajectory}. The pathological behavior associated with such a model error exists in many ad hoc data-driven statistical models for time series of partial observations of nature \cite{majda2012fundamental}. In addition, the PDF of $u_2$ in the imperfect model \eqref{eq: Triad_model_imperfect} is Gaussian by design, which is also different from the skewed PDF as in the perfect model.

\subsubsection{The experiment setup}
In this experiment, $u_1$ is the observed variable while $u_2$ and $u_3$ are unobserved in the triad model \eqref{eq: Triad_model}. The standard deviation of the observational noise is 0.2. The total length of the observational sequence is 16,000 (800 time units) and is separated into training and validation data set with a ratio 9:1. The testing set is the next 1,600 observational steps (from 800 to 880 time units). The LSTM model is trained based on 50 sampled trajectories of $u_1$, $u_2$, and $u_3$.

\subsubsection{The prediction skill}\label{sec: sampling_smoothing}

\begin{figure}[ht]
	\centering
	\includegraphics[width=\textwidth]{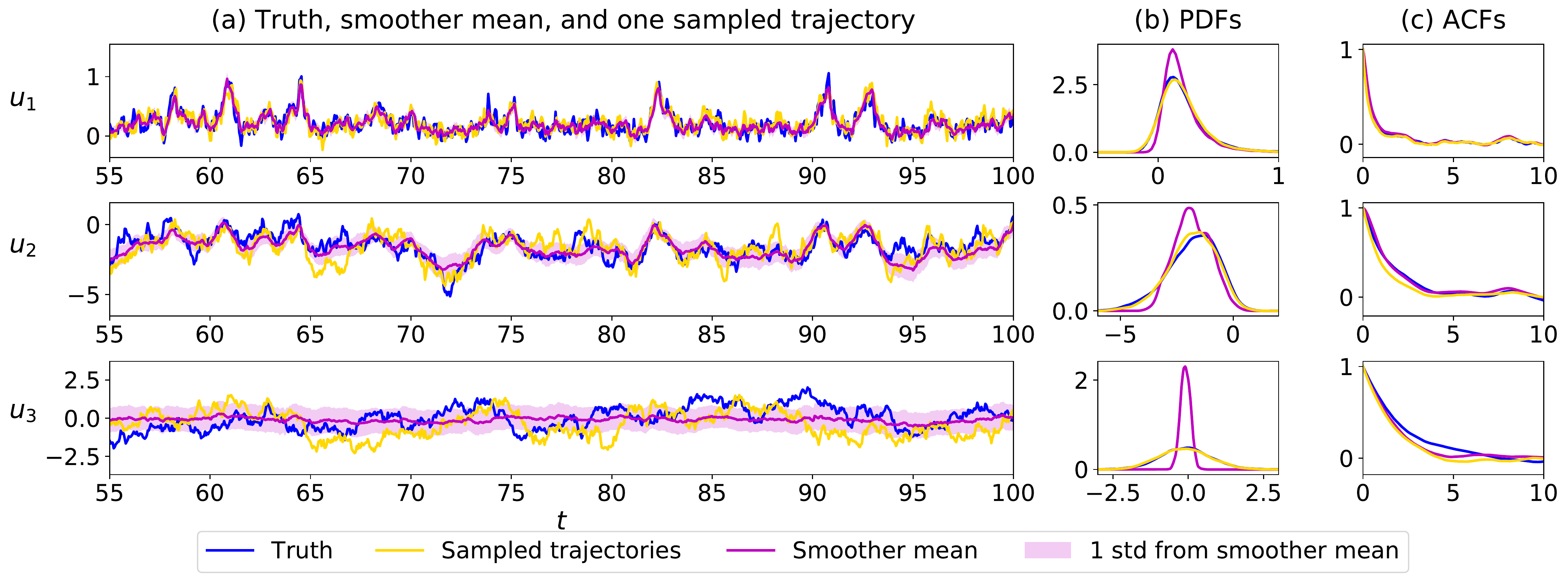}
	\caption{Comparison of the true signal from the triad model~\eqref{eq: Triad_model} with the smoother mean and the sampled trajectories based on the imperfect model \eqref{eq: Triad_model_imperfect}. Blue curves show the true trajectories, the associated PDFs and ACFs. The yellow curves show the posterior mean time series from the EnKS. The pink curves show one sampled trajectories. The uncertainty of the smoother estimates, represented by the one standard deviation, is shown by the pink shading areas. Panel (a): the time series of true signal, smoother mean, and one sampled trajectory. Panel (b)-(c): the PDFs and ACFs of the true signal, the smoother mean, and the sampled trajectory.}
	\label{fig: Triad_Sampling}
\end{figure}

Before discussing the prediction skill using the imperfect and the LSTM model, we start with showing the sampled trajectories and the posterior mean time series, which is the typical outcome from the traditional reanalysis techniques. The purpose here is to illustrate that the latter fails to capture the basic dynamical and statistical features in this tough test experiment. Therefore, the ML prediction can be biased and result in an inaccurate uncertainty quantification if it is trained on the posterior mean time series.
The smoother posterior mean time series (pink curves) are shown in Panel (a) of Figure~\ref{fig: Triad_Sampling}. It is clear that the dynamical behavior of the posterior mean time series is quite different from the true signals. As is shown in Panel (b), the large error in the PDFs, especially for $u_2$ and $u_3$, is as expected since the posterior mean time series smoothed out all the fluctuations that are also crucial for characterizing the underlying dynamics. In contrast, the sampled trajectories (yellow curves) resemble the truth in terms of both the dynamical and statistical features. Furthermore, they are very different from a free run of the imperfect model, as the observations serve as a regularizer in the Bayesian sampling process that completely resolved the instability problem that occurs in the imperfect model simulations.

\begin{figure}[ht]
	\centering
	\includegraphics[width=\textwidth]{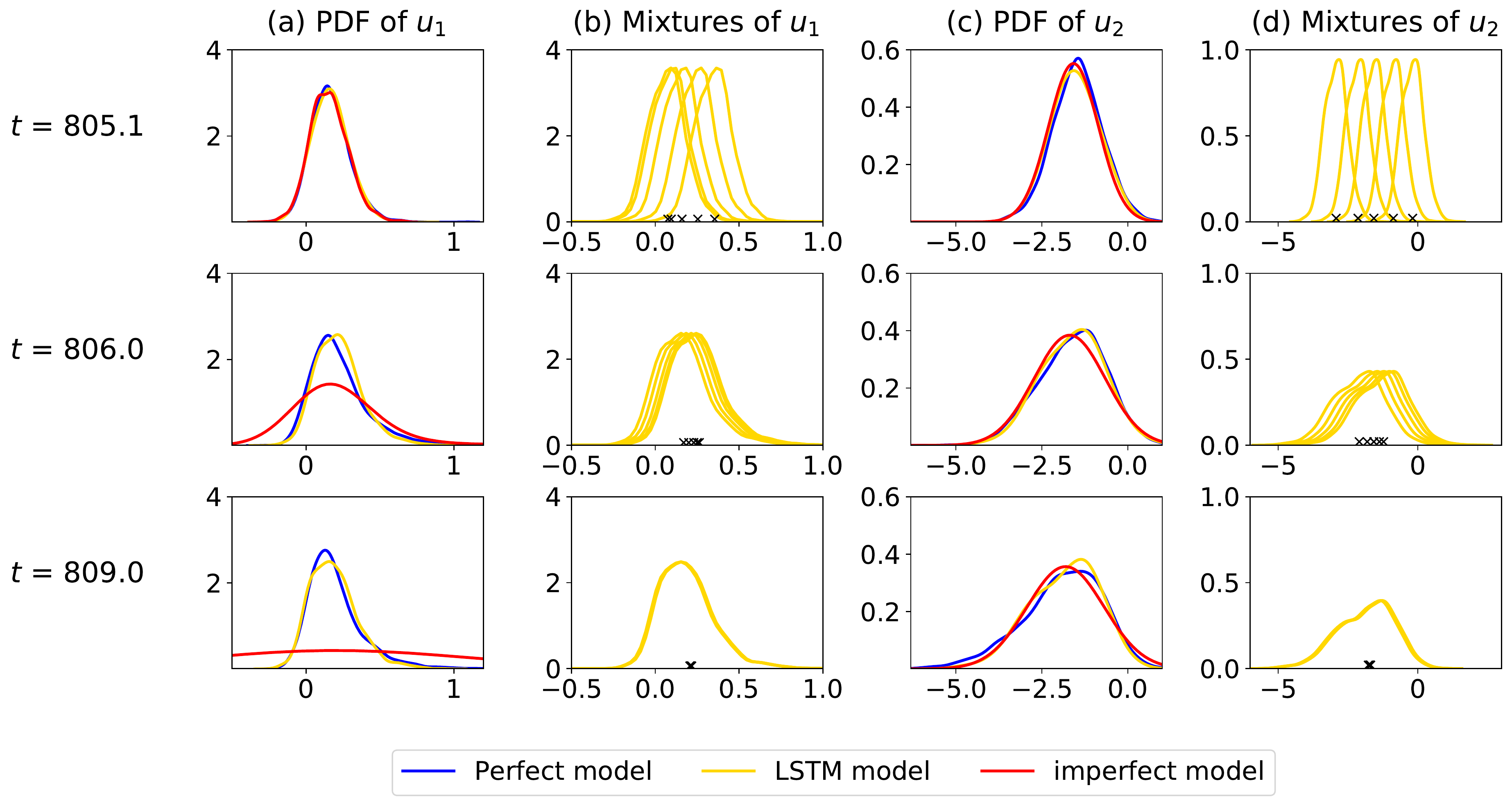}
	\caption{The time evolutions of the forecast uncertainty starting from $t=805$, where the true signal is generated from the triad model~\eqref{eq: Triad_model}. In Panel (a), the forecast PDFs of $u_1$ from the perfect model~\eqref{eq: Triad_model} and from the imperfect model~\eqref{eq: Triad_model_imperfect} are constructed using 2500 ensemble members whereas the PDFs from the LSTM model are constructed by 50 non-Gaussian mixture components where each component is formed by 50 points from the distribution~\eqref{eq: residual}. Panel (b) shows 5 out of the 50 components of the mixture distribution, corresponding to the 1st, 16th, 50th, 84th, 99th percentile of the point-wise forecast values i.e., $\hat{x}^k_{N+l}$ in \eqref{eq: mean}, among all the mixture components. The black cross-marks are these point-wise forecast values. Panel (c)--(d): similar to Panel (a)--(b) but for $u_2$. }
	\label{fig: triad_pdf}
\end{figure}

In Figure~\ref{fig: triad_pdf}, the time evolutions of the predicted PDFs for $u_1$ and $u_2$ starting from $t=805$ with the assimilated initial conditions are shown in Panels (a) and (c), respectively. For ensemble forecast using both the perfect and the imperfect parametric models, each PDF is constructed using 2500 ensemble members. The PDFs from the LSTM forecast model are constructed by 50 non-Gaussian mixture components where each component is formed by 50 points from the distribution~\eqref{eq: residual}. 
It is clear that, as time increases, the PDFs of $u_1$ constructed from the imperfect model become much more fat-tailed than those from the perfect model. This is because the nonlinear feedback~$-Iu_{1}^2$ is dropped in the imperfect model \eqref{eq: Triad_model_imperfect} such that the energy-conserving nonlinear constraint is broken. The consequence is that $u_1$ can stay in the unstable phase for a longer time, which triggers extremely large amplitudes of $u_1$ intermittently.
On the other hand, since $u_2$ is approximated by a linear Gaussian process in the imperfect model \eqref{eq: Triad_model_imperfect}, the PDF of $u_2$ of the imperfect model can never reflect any non-Gaussian information such as the skewness in the truth. In contrast, the PDFs constructed from the LSTM model are closer to the true PDFs since the training data is improved by the Bayesian sampling in the BAMCAFE algorithm. The skewness for the last row of $u_2$ from the perfect model, the imperfect model, and the LSTM model are -0.581, -0.035, -0.366, respectively. 

Panels (b) and (d) of Figure~\ref{fig: triad_pdf} show 5 out of the 50 components of the mixture distribution, corresponding to the 1st, 16th, 50th, 84th, 99th percentile of the point-wise forecast values i.e., $\hat{x}^k_{N+l}$ in \eqref{eq: mean}, among all the mixture components. The black cross marks are these point-wise forecast values. These mixture components show that the uncertainty of the prediction comes from two places. One is from the spread of all the mixture components. The other is from the LSTM model's uncertainty, quantified by the variability of each mixture component that is computed in light of the validation error \eqref{eq: residual}.
When the forecast lead time is small, the uncertainty of the LSTM model is tiny. In such a situation, the spread of the mixture components due to the use of DA for initialization is the dominant contribution to the uncertainty. On the other hand, as the forecast lead time increases, different mixture components gradually converge to each other and the uncertainty is explained  more and more by the single LSTM model's forecast error \eqref{eq: residual}.

\begin{figure}[ht]
	\centering
	\includegraphics[width=\textwidth]{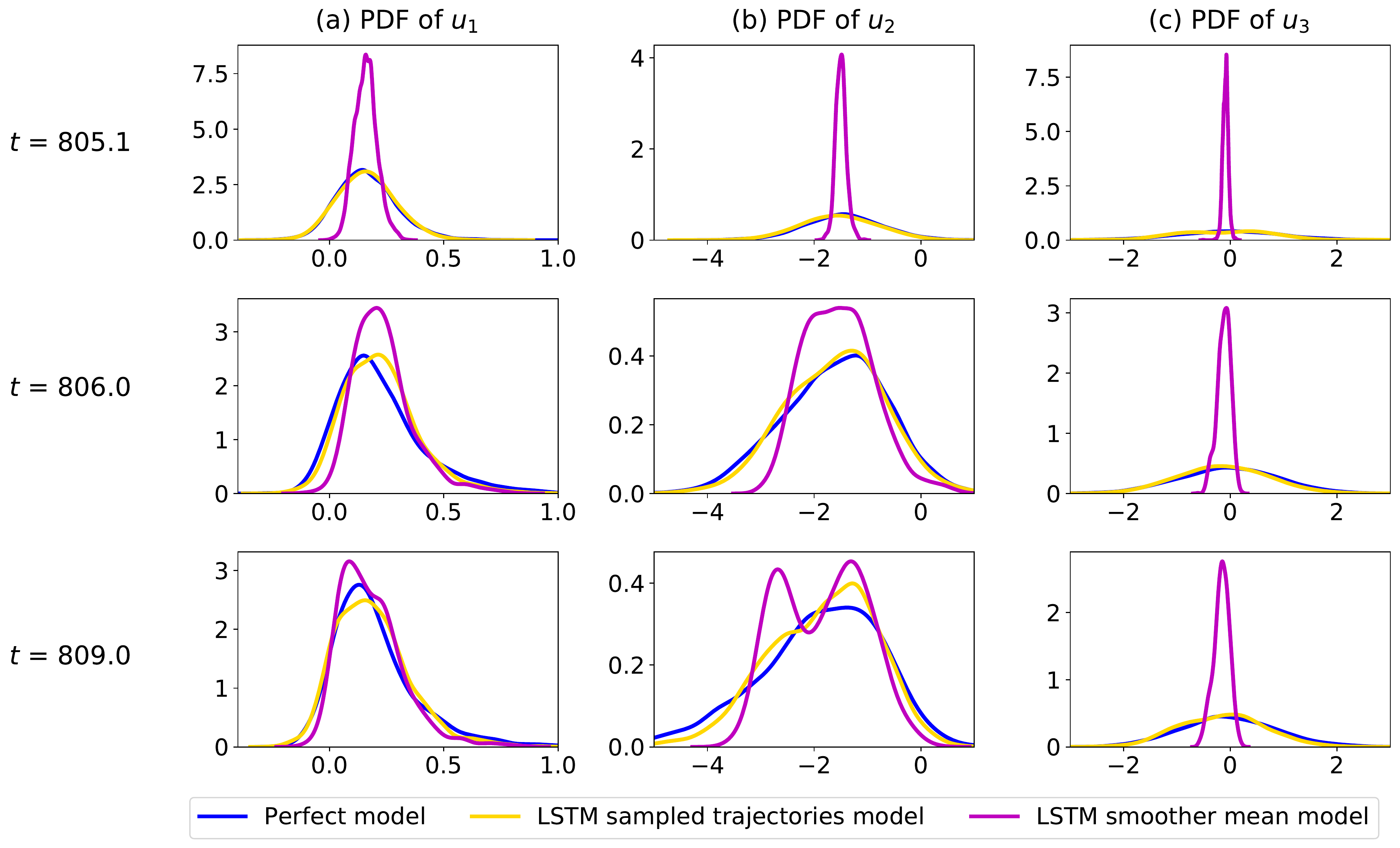}
	\caption{Comparison of the time evolutions of the forecast uncertainty starting from $t=805$ using different methods, where the true signal is generated from the triad model~\eqref{eq: Triad_model}. Panel (a)--(c): PDFs of $u_1$, $u_2$, and $u_3$, respectively. The blue curves are the forecasts using the perfect model~\eqref{eq: Triad_model}. The yellow curves are those using the LSTM model trained from sampled trajectories, as in Figure \ref{fig: triad_pdf}. The pink curves are the forecasts using the LSTM model but is trained from the smoother mean time series. }
	\label{fig: triad_pdf_w_smoother}
\end{figure}

Finally, since in traditional reanalysis, the smoother mean is usually used as a surrogate of the true signal, we compare the performance of the LSTM model trained from the sampled trajectories and the LSTM model trained from the smoother mean (generated from the same period of observations) in Figure~\ref{fig: triad_pdf_w_smoother}. It is clear that the PDFs constructed from the LSTM model trained from the smoother mean time series strongly underestimates the uncertainty. The fundamental reason is that the posterior mean time series has much less variability than the true signal, which has been shown in Figure \ref{fig: Triad_Sampling}. It is also worthwhile to note that the long-term forecast uncertainty  should be the same as the equilibrium PDF of the training time series. Yet, the pink curve of $u_2$ at $t=809$ in Figure \ref{fig: triad_pdf_w_smoother}, which corresponds to the forecast at the lead time of $4$ units that is much longer than the decorrelation time, does not equal to the equilibrium PDF of $u_2$ associated with the posterior mean in Figure \ref{fig: Triad_Sampling}. Such an error probably comes from the insufficient number of the data in the training set. In fact, if the posterior mean time series is used as training, then only one time series is contained in the training data set. In contrast, using multiple sampled trajectories effectively increases the size of the training data, which facilitate the training of the ML.

\clearpage

\section{Discussions and conclusion}\label{sec: discussion_conclusion}

A simple but effective Bayesian machine learning advanced forecast ensemble (BAMCAFE) method is developed that combines an available imperfect physics-informed model with DA to facilitate the ML ensemble forecast. In the BAMCAFE framework, a Bayesian ensemble DA is applied to create the training data of the ML model, which reduces the intrinsic error in the imperfect physics-informed model and provides the training data of the unobserved variables. Then a generalized DA is employed for the initialization of the ML ensemble forecast. In addition to providing the optimal point-wise forecast value, the BAMCAFE also improves the accuracy of quantifying the forecast uncertainty utilizing a non-Gaussian probability density function that characterizes the intermittency and extreme events. Furthermore, one advantage of using ML model to forecast is efficiency. For example, one can generate 2500-ensemble members of 5-time units forecast in about 30 minutes using the perfect L96 model on a server with a 64-core CPU, 256G memory. In contrast, it only takes a few seconds to generate the same forecast using new LSTM-based models.

Compared to the past studies that have already combined DA and ML, the novelties of this paper are as follows.
\begin{compactitem}
    \item[1.] An effective offline DA is applied to create an improved data set compared with the simulations from the available imperfect model. The new data set contains a recovery of the unobserved state variables. It also mitigates the model error. The augmented data is used as the training data for the ML models.
    \item[2.] Different from the traditional re-analysis technique, where only the optimal point-wise state estimation is often retained, the trajectories of all the ensemble members are utilized here as the ML training data. These sampled trajectories contain crucial temporal information of the underlying dynamics and the multiple trajectories facilitate a relatively short ML training period. 
    \item[3.] A generalized DA is developed for the ML forecast using only partial observations. 
    \item[4.] The forecast uncertainty is characterized in light of a non-Gaussian PDF via a mixture distribution.
\end{compactitem}

Future work includes the theoretical studies of the BAMCAFE algorithm, especially the quantification of the uncertainty, and applying the BAMCAFE algorithm to more complicated and realistic systems, such as those for numerical weather forecasting. As is expected, there will be some additional challenges of the latter. One such challenge is the high dimensionality of the system, which can affect the accuracy of the ensemble data assimilation results. Natural remedies include incorporating the localization and noise inflation techniques \cite{evensen2009data} into the ensemble data assimilation schemes or developing suitable approximate data assimilation algorithms that exploit semi-analytic formulae for the state estimation \cite{chen2018conditional}. Another difficulty for dealing with more realistic applications is the development of suitable approximate models. Skillful stochastic parameterizations and systematic model reduction strategies facilitate the application of the BAMCAFE. More sophisticated ML model architectures and training techniques, such as the mixed-scale dense convolutional neural network \cite{pelt2018mixed}, are also required for the BAMCAFE to forecast more efficiently and accurately.

\section*{Acknowledgments}
The research of N.C. is partially funded by the Office of VCRGE at UW-Madison and ONR N00014-21-1-2904.  Y. L. is supported as a graduate student under the these grants. The research of Y.L. was also supported in part by NSF Award DMS-2023239 through the IFDS at UW-Madison.

\section*{Data Availability Statement}
The source code for generating data and training LSTM models is openly available at~\url{https://github.com/zjkliyingda/BAMCAFE}.

\clearpage
\appendix
\section{Details about the ensemble Kalman filter and smoother}\label{sec: SI_EnKF}
Assume that the prior model from time $t_{n - 1}$ to $t_n$ is given as following form:
\begin{equation}\label{eq: forecast_model}
	\x^M_n = \cF^M(\x^M_{n - 1}) +  \bsigma^M_n.
\end{equation}
The goal of filtering problem is to estimate the unknown true state $\x_{n}$, given noisy and partial observations
\begin{equation}
	\y_n = \H\x_{n} +  \bsigma_n^o,
\end{equation}
where $\y_n$ is a $m$-dimensional measurement vector, $\H$ is a $m$ by $n$ matrix, and $ \bsigma_n^o$ is a vector of unbiased Gaussian noise with variance $\R^o_n$. Thus, the filtering problem can be formulated as maximizing the conditional distribution $p(\x_n | \y_{1:n})$ sequentially for each time $t = 1, 2, \cdots$ Based on the Bayesian's formula, the conditional distribution satisfies

\begin{equation}
	p(\x_n | \y_{1:n}) \propto p(\x_n | \y_{1:n -1}) p(\y_n | \x_n),
\end{equation}
where the prior estimates $p(\x_n | \y_{1:n - 1})$ is obtained from underlying model forecast~\eqref{eq: forecast_model} and $p(\y_n | \x_n)$ is the likelihood of estimating $\x_n$ given observation $\y_n$.

Ensemble Kalman filter (EnKF) is one of the typical methods used for filtering the nonlinear underlying dynamics. It is based on an assumption that the prior distribution is approximated by Gaussian and the mean and covariance are formed by a finite number $K$ of sampled members. More specifically, assume that the filtering ensemble at time $t_{n - 1}$ is given by $\{\x_{n - 1| n - 1}^{k}\}_{k=1, \ldots, K}$, the forecast ensemble $\{\x_{n| n - 1}^{k}\}_{k=1, \ldots, K}$ is obtained by applying forecast model~\eqref{eq: forecast_model} to each ensemble member. Thus, the prior distribution is
\begin{equation}
	p(\x_n | \y_{1:n - 1}) = \cN(\x_n | \bmu_{n|n - 1}, \bSigma_{n|n - 1}),
\end{equation}
where $\cN(\x| \bmu, \bSigma)$ denotes a multivariate normal distribution with mean $\bmu$ and covariance matrix $\bSigma$, and $\bmu_{n|n - 1}$ and $\bSigma_{n|n - 1}$ are sample mean and sample covariance, respectively. Given this Gaussian prior, the posterior distribution is also Gaussian
\begin{equation}
	\begin{aligned}
		p(\x_n | \y_{1:n}) \propto & ~ p(\x_n | \y_{1:n - 1}) p(\y_n | \x_n) \\
		= & ~\cN(\x_n | \bmu_{n|n - 1}, \bSigma_{n|n - 1}) \cN(\y_n | \H_n \x_n, \R^o_n) \\
		\propto & ~ \cN (\x_n | \bmu_{n|n}, \bSigma_{n|n}),
	\end{aligned}
\end{equation}
where the filtered mean $\bmu_{n|n}$ and filtered covariance $\bSigma_{n|n}$ are given by
\begin{equation}
	\begin{aligned}
		\bmu_{n|n} & = \bmu_{n|n - 1} + \K_n (\y_n - \H_n \bmu_{n|n - 1}), \\
		\bSigma_{n|n} & = (\I - \K_n \H_n) \bSigma_{n|n - 1},
	\end{aligned}
\end{equation}
with the Kalman gain $\K_n$
\begin{equation}
	\K_n = \bSigma_{n|n - 1} \H'_n (\H_n \bSigma_{n|n - 1}+ \R^o_n)^{-1}.
\end{equation}
The ensemble Kalman filter algorithm is given in Algorithm~\ref{alg: SI_EnKF}.

\begin{algorithm2e}[H]
	\caption{Ensemble Kalman filter}
	\label{alg: SI_EnKF}
	Start with an initial ensemble $\{\x^k_{0|0}\}_{k=1, \dots, K}$ sampled from initial distribution $\mathcal{N}(\bmu_{0|0}, \R_{0|0})$\;
	\For{$n = 1, \dots $}{
		\For{$k = 1, \dots, K$}{
			Compute forecast ensemble members using $\x^{k}_{n | n - 1} = \cF^M(\x^{k}_{n - 1}) +  \bsigma^M_n$\;
			Estimate prior ensemble covariance $\bSigma_{n|n-1}$ by the samples $\{\x^{k}_{n | n - 1}\}$\;
			Compute the Kalman gain $\K_{n} = \bSigma_{n|n-1} \mathbf{H}'_n(\mathbf{H}_n \bSigma_{n|n-1} \mathbf{H}'_n + \R^o_n)^{-1}$\;
			Draw a sample $\mathbf{v}^k_{n} \sim \mathcal{N}(0, \R^o_n)$ \;
			Update ensemble member $\x^{k}_{n | n} = \x^{k}_{n | n - 1} + \K_{n}(\x_n - \mathbf{H}_n \x^{k}_{n | n - 1} -\mathbf{v}^k_n)$ \;
		}
	}
\end{algorithm2e}

For smoothing, it exploits the information in the entire observational window, including the past and the future. Therefore, the smoothing problem optimizes the entire state $\{\x_{0:n}\}$ given all available observations $\{\y_1, \dots, \y_n\}$. Ensemble Kalman smoother (EnKS)~\cite{evensen2000ensemble} can be easily extended from EnKF.~It starts from the sequential formulation
\begin{equation}
	\begin{aligned}
		p(\x_{0:n} | \y_{1:n}) \propto & ~ p(\x_{0:n - 1} | \y_{1:n - 1})p(\x_n | \x_{n - 1}) p(\y_n | \x_n) \\
		= & ~ p(\x_{0:n} | \y_{1:n - 1}) p(\y_n | \x_n),
	\end{aligned}
\end{equation}
which is obtained using Bayesian's formula, the definition of the conditional PDF and the Markov property of the system. Thus, the vanilla version of EnKS algorithm is given by Algorithm~\ref{alg: SI_EnKS}.~\cite{evensen2000ensemble,katzfuss2020ensemble,cosme2012smoothing}

\begin{algorithm2e}[H]
	Start with an initial ensemble $\{\x^k_{0|0}\}_{k=1, \dots, K}$ sampled from initial distribution $\mathcal{N}(\mu_{0|0}, \R_{0|0})$\;
	\For{$n = 1, \dots $}{
		\For{$k = 1, \dots, K$}{
			Compute forecast ensemble members using $\x^{k}_{n | n - 1} = \cF^M(\x^{k}_{n - 1}) +  \bsigma_n$\;
			Estimate prior ensemble covariance $\bSigma_{n|n-1}$ by the samples $\{\x^{k}_{n | n - 1}\}$\;
			\For{$n' = 0, \dots, n$}{
				Compute cross-covariance $\bSigma_{n', n|n-1}$ between ensemble $\{\x^{k}_{n' | n - 1}\}$ and $\{\x^{k}_{n | n - 1}\}$ \;
				Compute the Kalman gain $\K_{n', n} = \bSigma_{n', n|n-1} \mathbf{H}'_n(\mathbf{H}_n \bSigma_{n,n|n-1} \mathbf{H}'_n + \R^o_n)^{-1}$\;
				Draw a sample $\mathbf{v}^k_{n} \sim \mathcal{N}(0, \R^o_n)$ \;
				Update ensemble member $\x^{k}_{n' | n} = \x^{k}_{n' | n - 1} + \K_{n', n}(\x_n - \mathbf{H}_n \x^{k}_{n | n - 1} -\mathbf{v}^k_n)$ \;
			}
		}
	}
	\caption{Ensemble Kalman smoother and sampling (EnKS)}\label{alg: SI_EnKS}
\end{algorithm2e}

\section{Parameters related to perfect model and data assimilation}
The parameters used in perfect models and data assimilation are shown in Table~\ref{tb: Parameters_perfect_DA}.

\begin{table}[!hbtp]
	\caption{Parameters used perfect model and data assimilation.}
	\label{tb: Parameters_perfect_DA}
\begin{tabularx}{\textwidth}{c *{14}{Y}}
	\toprule
	 & \multicolumn{12}{c}{model parameters}  
	 & \multicolumn{2}{c}{data assimilation}\\
	 \cmidrule(lr){2-13} \cmidrule(lr){14-15}
	L96 & $I~~$  & $J~~$ &  $h~~$ & $c~~$ & $b~~$ & $f~$ & $\sigma_{u_i} $ & $\sigma_{v_{i, j}} $ &  &  & & &  $\sigma^o$ & $\Delta_{\textrm{obs}} t$ \\
		& 40 & 4 & 2 & 2 & 2 & 4 & 1 & 1 &  & &  & & 1 & 0.05 \\
	\cmidrule(lr){2-13} \cmidrule(lr){14-15}
	Triad model   &  $\gamma_1$ & $\gamma_2$ & $\gamma_3$  &  $L_{12}$ & $L_{13}$ & $L_{23}$ & $I$ & $\delta$ & $\sigma_1$ & $\sigma_2$ & $\sigma_3$ & $F$ & $\sigma^o$ &  $\Delta_{\textrm{obs}} t$  \\
	& 2  & 0.2  & 0.4 & 0.2  & 0.1 & 0 & 5 & 1 & 0.5 & 1.2 & 0.8 & 2 & 0.2 & 0.05 \\
	\bottomrule
\end{tabularx}
\end{table}

\section{Hyperparameters used in the LSTM models}\label{sec: SI_Hyperparameters}

\begin{table}[!hbtp]
	\caption{Hyperparameters used in the LSTM models.}
	\label{tb: Hyperparameter_L96}
	\begin{tabularx}{\textwidth}{c * 6 {Y}}
	  \toprule
	  \midrule
	  L96 LSTM models \\
		 Lead time &  [1, 10]  & [11, 16] & [16, 70]  & [71, 100]\\
	  Maximum epochs &  300 &  100  & 100   & 100 \\
	  Hidden dimension & 64 & 64  &  32  &  16 \\
	  Learning rate &  \multicolumn{4}{c}{0.001}\\
	  Batch size & \multicolumn{4}{c}{64} \\
	  $L_{\textrm{init}}$ & \multicolumn{4}{c}{15 (0.75 time units)} \\
	  \midrule
	  \midrule
	  Triad LSTM model \\
	  Lead time &  [1, 3]  & [4, 15] & [16, 40] & [41, 80] \\
	  Maximum epochs &  200 &  100  & 30   & 30   \\
	  Hidden dimension & 64 & 64 & 32 & 16 \\
	  Learning rate &  \multicolumn{4}{c}{0.0005}\\
	  Batch size & \multicolumn{4}{c}{64} \\
	  $L_{\textrm{init}}$ & \multicolumn{4}{c}{15 (0.75 time units)} \\
	  \midrule
	  \midrule
	  Triad LSTM model (smoother mean) \\
	  Lead time &  [1, 3]  & [4, 15] & [16, 40] & [41, 80] \\
	  Maximum epochs &  1000 &  500  & 500   & 500   \\
	  Hidden dimension & 64 & 64 & 32 & 16 \\
	  Learning rate &  \multicolumn{4}{c}{0.001}\\
	  Batch size & \multicolumn{4}{c}{64} \\
	  $L_{\textrm{init}}$ & \multicolumn{4}{c}{15 (0.75 time units)} \\
	  \midrule
	  \bottomrule
	\end{tabularx}
  \end{table}

The hyperparameters used in the LSTM models associated with the two imperfect L96 models and the imperfect triad model are listed in Table~\ref{tb: Hyperparameter_L96}.

\bibliographystyle{plain}  
\bibliography{references}  

\end{document}